\pdfoutput=1
\documentclass[onecolumn]{sdl93}

\def\0{\mbox{\tiny $0$}}
\def\1{\mbox{\tiny $1$}}
\def\2{\mbox{\tiny $2$}}
\def\3{\mbox{\tiny $3$}}
\def\4{\mbox{\tiny $4$}}
\def\5{\mbox{\tiny $5$}}
\def\6{\mbox{\tiny $6$}}
\def\7{\mbox{\tiny $7$}}
\def\8{\mbox{\tiny $8$}}
\def\9{\mbox{\tiny $9$}}

\def\INC{\mathrm{INC}}
\def\REF{\mathrm{REF}}
\def\TRA{\mathrm{TRA}}

\def\LiSo{\mathrm{[LiSo]}}
\def\SoLiP{\mathrm{[SoLiP]}}
\def\SoLiS{\mathrm{[SoLiS]}}

\def\P{\mathrm{P}}
\def\S{\mathrm{S}}

\def\GH{\mathrm{GH}}

\logo{
\colorbox{DarkGoldenrod}{\color{white}$\mathbf{\Sigma\hspace*{0.06cm} \delta \hspace*{0.04cm} \Lambda}$}
}

\journal{\shadowtext{\textbf{\color{DarkRed} Pure and Applied Geophysics}} \,\,\,\textbf{175}, 2023-2044 (2018). }

\titlelines{3}
\title{Incidence angles maximizing\\ the Goos-H\"{a}nchen shift in\\ seismic data analysis}

\imgbgabstract{In the solid/liquid and liquid/solid scenarios, for the cases in which the P and S reflected waves are represented by  complex amplitudes, we give the closed formulas for the Goos-H\"anchen phase from which we can then  determine the lateral displacements. We compare the  results of the analysis done by using the Zoeppritz equations  with the calculations which appear in Optics.  We also discuss under which circumstances the plane wave analysis is valid and what happens for critical incidence where divergences appear. For the liquid/solid interface, the incidence angles maximizing the lateral displacement  are  given as solutions of  a polynomial equation.}

\author{
\names{Stefano De Leo$^{1,a}$ and Rita K. Kraus$^{2}$}
\affiliation{$^{1}$Department of Applied Mathematics, State of University of Campinas, Brazil.}
\affiliation{$^{2}$Institute of Geoscience, State University of Campinas, Brazil.}
\email{$^{a}$deleo@ime.unicamp.br}
}

\begin{document}

\sdlmaketitle

\section{Introduction}
Seismic reflection is a method of exploring the Earth's crust by artificially generated waves\cite{book1,book2,book3}. When the waves  vibrations meet the resistance of a different medium, they are reflected back as an echo. Geophysicists have mastered the technique of sound waves reflection and, using the fact that seismic amplitudes at each interface contain the physical properties just above and just below the boundary, employed reflection and transmission amplitudes in hydrocarbon detection, lithology identification, and fluid analysis. Seismic analysis essentially generates an ultrasonic image of the Earth. Special ships, fitted with equipment that produces seismic waves, are used to locate oil and gas. These waves fan out below the surface of the water, penetrating the seabed below and, depending on what they hit (rock, oil or gas) are reflected at different speeds back up towards the ship. There, seismometers record the waves and how fast they are travelling. These microphones are evenly spaced along a cable up to dozens of kilometers long, which is dragged behind the ship along the surface of the water. On land, vibrations are generated by vibrators, mounted on purpose-built vehicles.
The data recorded by the seismometers is logged by powerful mobile computers. Using this method, it is then possible to explore as deep as dozens of kilometers with remarkable accuracy.
Geophysicists process and analyze the collected data, with the help of computers, to generate detailed 3D models of the subsurface.

In geophysical exploration, we rarely deal with a simple isolated interface. However, our understanding of the variation of reflection and transmission coefficients with the incidence angle just begins with a single interface model. In this paper, we present a detailed analytical study of the Zoeppritz-Goos-H\"anchen (ZGH) shift for P waves in liquid/solid interfaces and for P/S waves in solid/liquid interfaces. The Maxwell-Goos-H\"anchen (MGH) shift is an optical phenomenon in which a light beam, when totally  reflected from an interface between two dielectric media having different refraction indices, undergoes a lateral shift from the position predicted by geometrical optics. The lateral shift, conjectured by Isaac Netwon in the 18th century, was experimentally observed, for the first time, by Fritz Gustav Goos and Hilda H\"anchen in 1947\cite{GH1} and, one year later, theoretically explained by Kurt Artmann\cite{ART} by using the stationary phase method. Due to the smallness of the shift, the MGH lateral displacement can be macroscopically  observed only by amplifying it by multiple reflections, as for example done in its first experimental detection, or using the technique based on the optical analog of the weak measurement in Quantum Mechanics\cite{WM1,WM2,WM3}, in which a combination of transverse electric  and magnetic waves generates an outgoing optical beam characterized by two peaks whose distance contains the information of the MGH shift.

Certainly, plane waves with their infinite wavefront do not represent physical beams as those used by geophysicists to collect data. Thus, calculations done by using plane waves are surely  useful to understand the problem but have to be  checked by a wave packet analysis. Nevertheless, the plane wave results often give the right behavior of physical beams.   For example,
most of the  analytical expressions obtained, in Optics,  for the lateral displacement are based on a plane wave analysis. However, it is important to observe that the plane wave approach is valid for incidence angle out of the critical region
\[\left[\,\arcsin\left(\frac{n_{\2}}{n_{\1}}\right)-\frac{\lambda}{\,{\mathrm{w}}_{\0}}\,\,\,,\,\,\,\arcsin\left(\frac{n_{\2}}{n_{\1}}\right)
+\frac{\lambda}{\,{\mathrm{w}}_{\0}}\, \right]\,\,,\]
where $\lambda$ is the wavelength of the optical beam, ${\mathrm{w}}_{\0}$ its minimal waist, and $n_{\1,\2}$ the refractive indices of the dielectrics above and below the boundary with $n_{\2}<n_{\1}$. In the critical region, a wave packet analysis is, indeed,  needed to remove the divergence found for incidence at the critical angle. In recent works\cite{ANA1,ANA2,ANA3,ANA4}, closed form expressions have been given in the critical region, where, due to the breaking of symmetry of the wave number distribution, an axial dependence is also present\cite{AXI1,AXI2}.

The considerable number of publications on this subject demonstrate an increasing interest, not only in the Optics community\cite{OPT1,OPT2} but also in other fields such as Particle Physics\cite{PP},  Condensed Matter\cite{CM}, and Geophysics\cite{GEO1,GEO2,GEO3,GEO4,GEO5}. In this paper, we shall concentrate our attention on the lateral shift of P/S waves reflected by a solid/liquid interface
and of P waves reflected by a liquid/solid interface because in these scenarios it is possible to obtain, by using the matrix form of the Zoeppritz equations, closed forms for the reflection and transmission coefficients and consequently  give analytical expressions for the Goos-H\"anchen phase. This allows to compare the Zoeppritz critical regions and the lateral displacements  with the ones given in Optics and determine when additional critical regions appear  and for which incidence angles maximal lateral displacements can be detected.

The main objective of this article is to provide seismologists with some of the fundamental mathematical background for understanding the role that Goos-H\"anchen  phase  plays in the sound wave propagation. With the increasing power and continuous evolution of computers, numerical calculations and data analysis can be used for improving the resolution and accuracy of seismic images. However, analytical formulas are fundamental for understanding the propagation properties of sound waves. In this spirit, closed form expressions for the ZGH phase allow to obtain polynomial equations from which the incidence angles maximizing the lateral displacement  can be calculated.
Starting from the  results  obtained  in this paper,  the authors hope is that  new investigations, developments of the  solutions, and possible applications could be stimulated in the Geophysics community.

The article is structured as follows. In Section II, we fix our notation and, to leave the presentation self-contained, give,  for an incident wave composed of a mixture of P and S waves, the unified matrix form of the Zoeppritz equations as well the energy flux partitioning at the interface. In Section III and IV, we respectively discuss the solid/liquid scenario for S and P waves, obtaining the analytical expressions for the reflection/transmission coefficients as well for the Goos-H\"anchen phase. Then, in Section V, we study
the liquid/solid scenario for an incident P wave.  Section VI contains the analysis of the lateral displacements
and, for the liquid/solid case, the polynomial equation from which  the incidence angles maximizing the ZGH shift can be determined. In this Section, we also compare the seismic with the optical results and discuss the regions of the incidence angles for which the results obtained by using the plane wave approach are valid.  For the graphical presentation of our results, the liquid chosen was water and the solids: vegetal soil, wet sand and granite. Conclusions, remarks and outlooks are given in the final Section.

\section{The matrix form of the Zoeppritz equations}
To calculate the reflection and transmission coefficients, we assume that the seismic P  and S waves can be treated as plane waves.  Discontinuity between the first and second elastic medium results in compressive as well as shearing stress. Then, arriving at the interface that separates the two elastic media,  both P and S waves are  reflected back and transmitted away from the interface.

Two classical methods for obtaining the plane waves reflection and transmission coefficients are often quoted in seismology textbooks. In 1899, Knott  gave the reflections and transmission coefficients by introducing unknown  potential amplitudes into the continuity conditions. In 1919, Zoeppritz found a very similar set of reflection and transmission coefficients by letting the unknown potential amplitudes be displacement amplitudes.  In this paper, we shall follow the notation and the matrix approach, based on the Zoeppritz method, given in the excellent book of Ikelle and Amundsen\cite{book2}  where the displacement amplitudes are replaced by the P and S wave potential amplitudes,  scaled by their respective velocities. In this formalism, the case of an incident P wave can be treated simultaneously with that of an incident S wave. Choosing $yz$  ($z$ being the axis perpendicular to the interface separating the medium 1 from the medium 2, see Fig.\,\ref{A}) as the plane of incidence, an incident  sound wave, containing a mixture $(\alpha,\beta)$ of P (represented by the potential amplitude $\Psi$)  and S  (represented by $\Phi$) waves and moving in medium 1, will be represented by
\begin{equation}
\alpha \,\,\Psi_{_{\INC}} + \beta \,\,\Phi_{_{\INC}}\,\,,
\end{equation}
with
\begin{equation}
\begin{array}{lcl}
\Psi_{_{\INC}} &=& v_{\1} \, \exp\left[\, i\,\omega \, \left(\, y\,\sin\psi_{\1} + z\, \cos\psi_{\1} - v_{\1}\,t \, \right) \,/\,v_{\1}\, \right]\\ & & \\
\Phi_{_{\INC}}&=& u_{\1} \, \exp\left[\, i\,\omega \, \left(\, y\,\sin\varphi_{\1} + z\, \cos\varphi_{\1} - u_{\1}\,t \, \right) \,/\,u_{\1}\, \right]\,\,,
\end{array}
\end{equation}
where $(v_{\1}\,,\,\psi_{\1})$  and $(u_{\1}\,,\,\varphi_{\1})$ are respectively the velocity and incidence angle of the P and S wave.  Reaching the interface, the incident wave will generate  reflected waves,
\begin{equation}
\begin{array}{lcl}
\Psi_{_{\REF}} & = & v_{\1} \, \left(\alpha\,R_{_{\P\P}}+\beta\,R_{_{\S\P}}\right) \, \exp\left[\, i\, \omega \, \left(\, y\,\sin\psi_{\1} - z\, \cos\psi_{\1} - v_{\1}\,t \, \right) \,/\, v_{\1}\, \right] \\ & & \\
\Phi_{_{\REF}} & = & u_{\1} \, \left(\alpha\,R_{_{\P\S}} + \beta\,R_{_{\S\S}}\right)\, \exp\left[\, i\,\omega\, \left(\, y\,\sin\varphi_{\1} - z\, \cos\varphi_{\1} - u_{\1}\,t \, \right) \,/\,u_{\1}\, \right]\,\,,
\end{array}
\end{equation}
moving back in medium 1, and transmitted waves
\begin{equation}
\begin{array}{lcl}
\Psi_{_{\TRA}} & = & v_{\2} \, \left(\alpha\,T_{_{\P\P}}+\beta\,T_{_{\S\P}}\right) \, \exp\left[\, i\, \omega \, \left(\, y\,\sin\psi_{\2} + z\, \cos\psi_{\2} - v_{\2}\,t \, \right) \,/\, v_{\2}\, \right] \\ & & \\
\Phi_{_{\TRA}} & = & u_{\2} \, \left(\alpha\,T_{_{\P\S}} + \beta\,T_{_{\S\S}}\right)\, \exp\left[\, i\,\omega\, \left(\, y\,\sin\varphi_{\2} + z\, \cos\varphi_{\2} - u_{\2}\,t \, \right) \,/\,u_{\2}\, \right]\,\,,
\end{array}
\end{equation}
moving forward in medium 2,  where $(v_{\2}\,,\,\psi_{\2})$  and $(u_{\2}\,,\,\varphi_{\2})$ are respectively the velocity and transmitted angle of the P and S wave.
The first subindex in the reflection and transmission coefficients indicates the nature of the incoming wave, $P$ or $S$. The second subindex instead represents the type of the reflected and  transmitted wave.  The minus sign in the phase of the reflected potentials indicate the propagation in the direction of the negative $z$ axis. The reflection and transmission angles are given by the Snell law
\begin{equation}
\frac{\sin\psi_{\1}}{v_{\1}} =\frac{\sin\varphi_{\1}}{u_{\1}} =\frac{\sin\psi_{\2}}{v_{\2}} =\frac{\sin\varphi_{\2}}{u_{\2}}\,\,.
\end{equation}
In Fig.\,\ref{A}, we show a schematic representation of incident, reflected and transmitted acoustic waves for granite/water and water/granite interfaces.

Due to the fact that the velocities are known and the angles are determined by the Snell law, the only quantities to be calculated are the reflection and transmission coefficients. These coefficients are obtained by imposing the continuity of displacement and stress across the interface.  In terms of wave potentials, this means the continuity of the following functions
\begin{equation}
1)\,\,\Psi_y -\Phi_z\,\,,\,\,\,\,\,
2)\,\,\Psi_z +\Phi_y\,\,,\,\,\,\,\,
3)\,\,\mu\,(\,2\,\Psi_{yz}+\Phi_{yy}-\Phi_{zz}\,)\,\,,\,\,\,\,\,
4)\,\,\lambda\,(\,\Psi_{yy}+\Psi_{zz}\,) + 2\,\mu\,(\,\Psi_{zz}+\Phi_{yz}\,)\,\,,
\end{equation}
where $\mu =\rho\,u^{\2}$ and $\lambda=\rho\,(\,v^{\2}-2\,u^{\2}\,)$, with $\rho$ the medium density and u (v) the velocity of the S (P) wave. After simple algebraic manipulations, we can rewrite the equations coming from the continuity of displacement and stress in the following compact matrix form
\begin{equation}
\label{matuni}
\mathcal{M}\, A = B\,\,,
\end{equation}
where
\[
\mathcal{M} = \left(\,
\begin{array}{rrrr}
-\,\sin\psi_{\1} & -\,\cos\varphi_{\1} &\sin\psi_{\2} &-\,\cos\varphi_{\2}\\
\cos\psi_{\1} & -\,\sin\varphi_{\1} & \cos\psi_{\2} & \sin\varphi_{\2} \\
\rho_{\1}u_{\1}^{\2}v_{\2}\sin2\psi_{\1} &  \rho_{\1}u_{\1}v_{\1}v_{\2}\cos2\varphi_{\1}& \rho_{\2}u_{\2}^{\2}v_{\1}\sin2\psi_{\2} & -\,\rho_{\2}u_{\2}v_{\2}v_{\1}\cos2\varphi_{\2}\\
-\,\rho_{\1}v_{\1}\cos2\varphi_{\1} & \rho_{\1}u_{\1}\sin2\varphi_{\1} & \rho_{\2}v_{\2}\cos2\varphi_{\2} & \rho_{\2}u_{\2}\sin2\varphi_{\2}
\end{array}
\,\right)
\]
and
\[
A=\left(\,\begin{array}{c}
\alpha\,R_{_{\P\P}}+\,\beta\,R_{_{\S\P}}\\
\alpha\, R_{_{\P\S}}+\,\beta\,R_{_{\S\S}}\\
\alpha\,T_{_{\P\P}}+\,\beta\,T_{_{\S\P}}\\
\alpha\, T_{_{\P\S}}+\,\beta\,T_{_{\S\S}}
\end{array}
\right)\,\,,\,\,\,\,\,\,\,B=\left[\,\begin{array}{c}
\alpha\,\sin\psi_{\1} \,-\,\beta\,\cos\varphi_{\1}\\
\alpha\,\cos\psi_{\1} \,+\,\beta\,\sin\varphi_{\1}\\
\rho_{\1}u_{\1}v_{\2}(\,\alpha\,u_{\1}\sin2\psi_{\1}\,-\,\beta\,v_{\1}\cos2\varphi_{\1})\\
\rho_{\1}(\,\alpha\,v_{\1}\cos 2\varphi_{\1}\,+\,\beta\,u_{\1}\sin 2\varphi_{\1})
\end{array}
\right]\,\,.
\]
The energy flux partitioning at the interface can be expressed in terms of the reflection and transmission coefficients appearing in Eq.\,(\ref{matuni}) as follows
\begin{equation}
\label{Keq}
\begin{array}{ccl}
E_{_{\REF(\P)}} & = &      \alpha\,\left|\,R_{_{\P\P}}\,\right|^{^{2}} +\,\beta\,\displaystyle{\frac{v_{\1}\cos\psi_{\1}}{u_{\1}\cos\varphi_{\1}}}\,\left|\,R_{_{\S\P}}\,\right|^{^{2}}\,\,,    \\ \\
E_{_{\REF(\S)}} & = & \alpha\,\displaystyle{\frac{u_{\1}\cos\varphi_{\1}}{v_{\1}\cos\psi_{\1}}}\,\left|\,R_{_{\P\S}}\,\right|^{^{2}} +\,\beta\,\left|\,R_{_{\S\S}}\,\right|^{^{2}}\,\,,     \\ \\
E_{_{\TRA(\P)}} & = &
\alpha\,\displaystyle{\frac{\rho_{\2}v_{\2}\cos\psi_{\2}}{\rho_{\1}v_{\1}\cos\psi_{\1}}}\,\left|\,T_{_{\P\P}}\,\right|^{^{2}}
+\,\beta\,\displaystyle{\frac{\rho_{\2}v_{\2}\cos\psi_{\2}}{\rho_{\1}u_{\1}\cos\varphi_{\1}}}\,\left|\,T_{_{\S\P}}\,\right|^{^{2}}\,\,, \\ \\
E_{_{\TRA(\S)}} & = &
\alpha\,\displaystyle{\frac{\rho_{\2}u_{\2}\cos\varphi_{\2}}{\rho_{\1}v_{\1}\cos\psi_{\1}}}\,\left|\,T_{_{\P\S}}\,\right|^{^{2}}
+\,\beta\,\displaystyle{\frac{\rho_{\2}u_{\2}\cos\varphi_{\2}}{\rho_{\1}u_{\1}\cos\varphi_{\1}}}\,
\left|\,T_{_{\S\S}}\,\right|^{^{2}}\,\,. \end{array}
\end{equation}
The Knott energy coefficients, given in Eq.\,(\ref{Keq}), satisfy the energy conservation
\begin{equation}
E_{_{\REF(\P)}} + E_{_{\REF(\S)}} + E_{_{\TRA(\P)}}  + E_{_{\TRA(\S)}}  =\,\, \alpha\, +\, \beta\,\,.
\end{equation}
In the following Sections, we discuss the solid/liquid and liquid/solid scenarios and, once  obtained the explicit solutions for the reflection and transmission coefficients, we give, for each case in which a critical region appears, the Goos-H\"anchen phase, responsible for  the lateral displacement of the reflected waves, and then discuss its properties. The graphical presentation of the results will be done  by using the media, velocities and densities listed in Table 1.

\section{Solid-liquid interface: incident S waves}

In the solid/liquid scenario, we can only have transmitted $P$ waves ($u_{\2}=\varphi_{\2}=0$). For the case of  incident S wave ($\alpha=0$ and $\beta=1$), the Knott energy coefficients become

\begin{equation*}
\{\,E_{_{\REF(\P)}}\,,\,E_{_{\REF(\S)}}\,,\,E_{_{\TRA(\P)}} \,,\,E_{_{\TRA(\S)}}\,\} =
\left\{\,\displaystyle{\frac{v_{\1}\cos\psi_{\1}}{u_{\1}\cos\varphi_{\1}}}\,\left|\,R_{_{\S\P}}\,\right|^{^{2}}\,,\,    \left|\,R_{_{\S\S}}\,\right|^{^{2}}\,,\,
\displaystyle{\frac{\rho_{\2}v_{\2}\cos\psi_{\2}}{\rho_{\1}u_{\1}\cos\varphi_{\1}}}\,\left|\,T_{_{\S\P}}\,\right|^{^{2}}\,
,\,0\,\right\}
\end{equation*}
and the reflection and transmission coefficients are obtained by solving the reduced Zoeppritz matrix equation

\begin{equation}
\left(\,
\begin{array}{rrrr}
-\,\sin\psi_{\1} & -\,\cos\varphi_{\1} &\sin\psi_{\2} &-\,1\\
\cos\psi_{\1} & -\,\sin\varphi_{\1} & \cos\psi_{\2} & 0 \\
u_{\1}\sin2\psi_{\1} & v_{\1}\cos2\varphi_{\1}& 0 & 0\\
-\,\rho_{\1}v_{\1}\cos2\varphi_{\1} & \rho_{\1}u_{\1}\sin2\varphi_{\1} & \rho_{\2}v_{\2} & 0
\end{array}
\,\right)\,\,\left(\,\begin{array}{c}
R_{_{\S\P}}\\
R_{_{\S\S}}\\
T_{_{\S\P}}\\
T_{_{\S\S}}
\end{array}
\right) =  \left(\,\begin{array}{c}
-\,\cos\varphi_{\1}\\
\sin\varphi_{\1}\\
-\,v_{\1}\cos2\psi_{\1}\\
\rho_{\1}u_{\1}\sin2\varphi_{\1}
\end{array}
\right)\,\,.
\end{equation}
For an S wave incident  upon a solid/liquid interface, we have four possible regions for the incidence angle
$\varphi_{\1}$: the region without critical angles,
\begin{equation}
0\,<\,\varphi_{\1}^{^{[\mathrm{a}]}}\,<\, {\mathrm{min}}\left[\,\arcsin\frac{u_{\1}}{v_{\1}}\,,\, \arcsin\frac{u_{\1}}{v_{\2}}\,\right]\,\,,
\end{equation}
the region with evanescent transmitted P waves (and travelling reflected P and S waves),
\begin{equation}
\arcsin\frac{u_{\1}}{v_{\2}}\,<\,\varphi_{\1}^{^{[\mathrm{b}]}}\,<\, \arcsin\frac{u_{\1}}{v_{\1}}\,\,,
\end{equation}
the region with evanescent reflected P waves (and travelling transmitted P and reflected S waves),
 \begin{equation}
\arcsin\frac{u_{\1}}{v_{\1}}\,<\,\varphi_{\1}^{^{[\mathrm{c}]}}\,<\, \arcsin\frac{u_{\1}}{v_{\2}}\,\,,
\end{equation}
and finally the region with evanescent reflected and transmitted P waves (and travelling reflected S waves),
\begin{equation}
 {\mathrm{max}}\left[\,\arcsin\frac{u_{\1}}{v_{\1}}\,,\, \arcsin\frac{u_{\1}}{v_{\2}}\,\right]\,<\,
 \varphi_{\1}^{^{[\mathrm{d}]}}\,<\,\frac{\pi}{2}\,\,.
\end{equation}
The critical angles for an incident S wave in the solid/liquid examples used in this paper are given in Table 2.

\subsection{Incidence region before the critical angles}
In this incidence region, $\cos\psi_{\1}$ and $\cos\psi_{\2}$ are both reals and we have two (P and S waves) reflected waves and a transmitted P wave. The Knott coefficients are given by
\begin{eqnarray}
E^{^{\SoLiS}}_{_{\REF(\P)}} & = &  \frac{v_{\1}\cos\psi_{\1}}{u_{\1}\,\cos\varphi_{\1}}\,\left[\,
\frac{2\,\rho_{\1}u_{\1}v_{\1}\sin2\varphi_{\1}\cos2\varphi_{\1}\cos\psi_{\2}}
{\rho_{\2} v_{\2}v_{\1}\cos\psi_{\1} + \rho_{\1}\cos\psi_{\2}\,(\,u_{\1}^{\2}\sin2\psi_{\1}\sin2\varphi_{\1} + v_{\1}^{\2}\cos^{\2}2\varphi_{\1}\,)}
\,\right]^{^{2}}\,\,, \nonumber \\
E^{^{\SoLiS}}_{_{\REF(\S)}} & = & \left[\,
\frac{\rho_{\2} v_{\2} v_{\1}\cos\psi_{\1} + \rho_{\1}\cos\psi_{\2}\,(\,-\, u_{\1}^{\2}\sin2\psi_{\1}\sin2\varphi_{\1} +  v_{\1}^{\2}\cos^{\2}2\varphi_{\1}\,)}
 {\rho_{\2} v_{\2} v_{\1}\cos\psi_{\1} + \rho_{\1}\cos\psi_{\2}\,(\,u_{\1}^{\2}\sin2\psi_{\1}\sin2\varphi_{\1} + v_{\1}^{\2}\cos^{\2}2\varphi_{\1}\,)}
\,\right]^{^{2}}\,\,,   \\
E^{^{\SoLiS}}_{_{\TRA(\P)}} & = & \frac{\rho_{\2}v_{\2}\cos\psi_{\2}}{\rho_{\1}u_{\1}\cos\varphi_{\1}}\,\left[\,
\frac{2\,\rho_{\1}u_{\1}v_{\1}\cos\psi_{\1}\sin2\varphi_{\1}}
{\rho_{\2} v_{\2} v_{\1}\cos\psi_{\1}+ \rho_{\1}\cos\psi_{\2}\,(\,u_{\1}^{\2}\sin2\psi_{\1}\sin2\varphi_{\1} + v_{\1}^{\2}\cos^{\2}2\varphi_{\1}\,)}
\,\right]^{^{2}}\,\,,   \nonumber
\end{eqnarray}
and satisfy the energy conservation equation
\[
E^{^{\SoLiS}}_{_{\REF(\P)}} + E^{^{\SoLiS}}_{_{\REF(\S)}} + E^{^{\SoLiS}}_{_{\TRA(\P)}} =1\,\,.
\]
The upper limit of the incidence angles which guarantees real reflection and transmission coefficients and consequently 3 travelling waves with velocities $u_{\1}$ (for S wave reflected in the first medium),  $v_{\1}$ (for P wave reflected in the first medium),   $v_{\2}$ (for P wave transmitted in the second medium),   is  then given by $11.54^{^{o}}$ for vegetal soil/water,   $17.46^{^{o}}$ for wet sand/water, and   $33.37^{^{o}}$ for granite/water,
see Fig.\,\ref{B}.

\subsection{Incidence region with evanescent transmitted P waves}
This is the zone of the incidence region for which $u_{\1}/v_{\2}<\sin\varphi_{\1}<u_{\1}/v_{\1}$.  In this case,
\[\cos\psi_{\2}= \sqrt{1 - \left(\frac{v_{\2}}{u_{\1}}\,\sin\varphi_{\1}\right)^{^{2}}} = i\, \sqrt{\left(\frac{v_{\2}}{u_{\1}}\,\sin\varphi_{\1}\right)^{^{2}}-1}=
 i\,|\cos\psi_{\2}|\,\,,
\]
the Knott coefficients become
\begin{equation}
\begin{array}{lcl}
E^{^{\SoLiS}}_{_{\REF(\P)}} & = &  \displaystyle{\frac{v_{\1}\cos\psi_{\1}}{u_{\1}\,\cos\varphi_{\1}}\,\left|\,
\frac{2\,i\,\rho_{\1}u_{\1}v_{\1}\sin2\varphi_{\1}\cos2\varphi_{\1}|\cos\psi_{\2}|}
{\rho_{\2} v_{\2}v_{\1}\cos\psi_{\1} + i\,\rho_{\1}|\cos\psi_{\2}|\,(\,u_{\1}^{\2}\sin2\psi_{\1}\sin2\varphi_{\1} + v_{\1}^{\2}\cos^{\2}2\varphi_{\1}\,)}
\,\right|^{^{2}}}\,\,,\\
E^{^{\SoLiS}}_{_{\REF(\S)}} & = & \displaystyle{\left|\,
\frac{\rho_{\2} v_{\2} v_{\1}\cos\psi_{\1} + i\,\rho_{\1}|\cos\psi_{\2}|\,(\,-\, u_{\1}^{\2}\sin2\psi_{\1}\sin2\varphi_{\1} +  v_{\1}^{\2}\cos^{\2}2\varphi_{\1}\,)}
 {\rho_{\2} v_{\2} v_{\1}\cos\psi_{\1} + i\,\rho_{\1}|\cos\psi_{\2}|\,(\,u_{\1}^{\2}\sin2\psi_{\1}\sin2\varphi_{\1} + v_{\1}^{\2}\cos^{\2}2\varphi_{\1}\,)}
\,\right|^{^{2}}}\,\,,
\end{array}
\end{equation}
and satisfy the energy conservation
\[
E^{^{\SoLiS}}_{_{\REF(\P)}} + E^{^{\SoLiS}}_{_{\REF(\S)}}  =1\,\,.
\]
This scenario is for example reproduced in the presence of the vegetal soil/water interface for incidence angles
between in $11.54^{^{o}}$ and $25.38^{^{o}}$, as illustrated  in Fig.\,\ref{B}(a-b).

In this incidence region, both the P and S reflected waves gain an additional GH phase. For the P waves, the ZGH phase is
\begin{equation}
\begin{array}{rcl}
\alpha_{_{\GH}}^{^{\SoLiS}} &=& -\,\displaystyle{\arctan\left[\,   \frac{
\rho_{\1}|\cos\psi_{\2}|\,(\,u_{\1}^{\2}\sin2\psi_{\1}\sin2\varphi_{\1} + v_{\1}^{\2}\cos^{\2}2\varphi_{\1}\,)}{\rho_{\2} v_{\2} v_{\1}\cos\psi_{\1}}\, \right]\,+\,\frac{\pi}{2}}\,\,,
\end{array}
\end{equation}
and, for the S waves,
\begin{equation}
\begin{array}{rcl}
\beta_{_{\GH}}^{^{\SoLiS}} &=& \displaystyle{\arctan\left[\,   \frac{
\rho_{\1}|\cos\psi_{\2}|\,(\,-\,u_{\1}^{\2}\sin2\psi_{\1}\sin2\varphi_{\1} + v_{\1}^{\2}\cos^{\2}2\varphi_{\1}\,)}{\rho_{\2} v_{\2} v_{\1}\cos\psi_{\1}}\, \right]} \,-\\ \\
 & & \displaystyle{\arctan\left[\,   \frac{
\rho_{\1}|\cos\psi_{\2}|\,(\,u_{\1}^{\2}\sin2\psi_{\1}\sin2\varphi_{\1} + v_{\1}^{\2}\cos^{\2}2\varphi_{\1}\,)}{\rho_{\2} v_{\2} v_{\1}\cos\psi_{\1}}\, \right]}\,\,.
\end{array}
\end{equation}

\subsection{Incidence region with evanescent reflected P waves}
For incidence angles in this region ($u_{\1}/v_{\1}<\sin\varphi_{\1}<u_{\1}/v_{\2}$)
$\cos\psi_{\1}=i\,|\cos\psi_{\1}|$ and the Knott coefficients are given by
\begin{equation}
\begin{array}{lcl}
E^{^{\SoLiS}}_{_{\REF(\S)}} & = & \displaystyle{\left|\,
\frac{i\,\rho_{\2} v_{\2} v_{\1}|\cos\psi_{\1}| + \rho_{\1}\cos\psi_{\2}\,(\,-\,i\, u_{\1}^{\2}|\sin2\psi_{\1}|\sin2\varphi_{\1} +  v_{\1}^{\2}\cos^{\2}2\varphi_{\1}\,)}
 {i\,\rho_{\2} v_{\2} v_{\1}|\cos\psi_{\1}| + \rho_{\1}\cos\psi_{\2}\,(\,i\,u_{\1}^{\2}|\sin2\psi_{\1}|\sin2\varphi_{\1} + v_{\1}^{\2}\cos^{\2}2\varphi_{\1}\,)}
\,\right|^{^{2}}}\,\,,   \\
E^{^{\SoLiS}}_{_{\TRA(\P)}} & = & \displaystyle{\frac{\rho_{\2}v_{\2}\cos\psi_{\2}}{\rho_{\1}u_{\1}\cos\varphi_{\1}}\,\left|\,
\frac{2\,i\,\rho_{\1}u_{\1}v_{\1}|\cos\psi_{\1}|\sin2\varphi_{\1}}
{i\,\rho_{\2} v_{\2} v_{\1}|\cos\psi_{\1}|+ \rho_{\1}\cos\psi_{\2}\,(\,i\,u_{\1}^{\2}|\sin2\psi_{\1}|\sin2\varphi_{\1} + v_{\1}^{\2}\cos^{\2}2\varphi_{\1}\,)}
\,\right|^{^{2}}}\,\,.
\end{array}
\end{equation}
The energy conservation is now guaranteed by
\[
 E^{^{\SoLiS}}_{_{\REF(\S)}} + E^{^{\SoLiS}}_{_{\TRA(\P)}} =1\,\,.
\]
This is for example the case of the wet sand/water interface for incidence angles in between $17.46^{^{o}}$ and $23.58^{^{o}}$, see Fig.\,\ref{B}(c-d), and of the  granite/water interface for incidence angles greater than $33.37^{^{o}}$, see Fig.\,\ref{B}(e-f).

In this incidence region, for the S reflected we have the following ZGH phase

\begin{equation}
\begin{array}{rcl}
\gamma_{_{\GH}}^{^{\SoLiS}} &=& \displaystyle{\arctan\left[\,   \frac{
\rho_{\2}v_{\2}v_{\1}|\cos\psi_{\1}|\,-\, \rho_{\1}u_{\1}^{\2}\cos\psi_{\2}|\sin2\psi_{\1}|\sin2\varphi_{\1}
 }{\rho_{\1}v_{\1}^{\2}\cos\psi_{\2}\cos^{\2}2\varphi_{\1}}  \, \right]}\, -\\ \\
 & &\displaystyle{\arctan\left[\,   \frac{
\rho_{\2}v_{\2}v_{\1}|\cos\psi_{\1}|\,+\, \rho_{\1}u_{\1}^{\2}\cos\psi_{\2}|\sin2\psi_{\1}|\sin2\varphi_{\1}
 }{\rho_{\1}v_{\1}^{\2}\cos\psi_{\2} \cos^{\2}2\varphi_{\1}}  \, \right]} \,\,.
\end{array}
\end{equation}

\subsection{Incidence region with evanescent reflected and transmitted P waves}
In this incidence region both $\cos\psi_{\1}$ and $\cos\psi_{\2}$ are imaginary and the Knott coefficient becomes
\begin{equation}
E^{^{\SoLiS}}_{_{\REF(\S)}}  =  \left|\,
\frac{i\,\rho_{\2} v_{\2} v_{\1}|\cos\psi_{\1}| + i\,\rho_{\1}|\cos\psi_{\2}|\,(\,-\,i\, u_{\1}^{\2}|\sin2\psi_{\1}|\sin2\varphi_{\1} +  v_{\1}^{\2}\cos^{\2}2\varphi_{\1}\,)}
 {i\,\rho_{\2} v_{\2} v_{\1}|\cos\psi_{\1}| + i\,\rho_{\1}|\cos\psi_{\2}|\,(\,i\,u_{\1}^{\2}|\sin2\psi_{\1}|\sin2\varphi_{\1} + v_{\1}^{\2}\cos^{\2}2\varphi_{\1}\,)}
\,\right|^{^{2}}=1\,\,.
\end{equation}
This is the scenario illustrated in the plots of Fig.\,\ref{B}  for vegetal soil/water (a-b) and wet sand/water (c-d) for incidence angles respectively greater than $25.38^{^{o}}$
and $23.58^{^{o}}$.

The reflection coefficient for the S wave is given by $\exp\{i\,\delta_{_{\GH}}^{^{\SoLiS}}\}$ with
\begin{equation}
\begin{array}{rcl}
\delta_{_{\GH}}^{^{\SoLiS}} &=& -\,2\,\displaystyle{\arctan\left[\,   \frac{ \rho_{\1}u_{\1}^{\2}|\cos\psi_{\2}\sin2\psi_{\1}|\sin2\varphi_{\1}}{
\rho_{\2}v_{\2}v_{\1}|\cos\psi_{\1}|\,+\, \rho_{\1}v_{\1}^{\2}|\cos\psi_{\2}|\cos^{\2}2\varphi_{\1}}  \, \right]}
\,\,.
\end{array}
\end{equation}

\section{Solid-liquid interface: incident P waves}
In the case of incident P waves ($\alpha=1$ and $\beta=0$) the Knott coefficients become
\begin{equation*}
\{\,E_{_{\REF(\P)}}\,,\,E_{_{\REF(\S)}}\,,\,E_{_{\TRA(\P)}} \,,\,E_{_{\TRA(\S)}}\,\} =
\left\{\,\displaystyle{\left|\,R_{_{\P\P}}\,\right|^{^{2}}\,,\,  \frac{u_{\1}\cos\varphi_{\1}}{v_{\1}\cos\psi_{\1}}} \, \left|\,R_{_{\P\S}}\,\right|^{^{2}}\,,\,
\displaystyle{\frac{\rho_{\2}v_{\2}\cos\psi_{\2}}{\rho_{\1}v_{\1}\cos\psi_{\1}}}\,\left|\,T_{_{\P\P}}\,\right|^{^{2}}\,
,\,0\,\right\}
\end{equation*}
and the  Zoeppritz matrix equation (\ref{matuni}) reduces to
\begin{equation}
\left(\,
\begin{array}{rrrr}
-\,\sin\psi_{\1} & -\,\cos\varphi_{\1} &\sin\psi_{\2} &-\,1\\
\cos\psi_{\1} & -\,\sin\varphi_{\1} & \cos\psi_{\2} & 0 \\
u_{\1}\sin2\psi_{\1} & v_{\1}\cos2\varphi_{\1}& 0 & 0\\
-\,\rho_{\1}v_{\1}\cos2\varphi_{\1} & \rho_{\1}u_{\1}\sin2\varphi_{\1} & \rho_{\2}v_{\2} & 0
\end{array}
\,\right)\,\,\left(\,\begin{array}{c}
R_{_{\P\P}}\\
R_{_{\P\S}}\\
T_{_{\P\P}}\\
T_{_{\P\S}}
\end{array}
\right) =  \left(\,\begin{array}{c}
\sin\psi_{\1}\\
\cos\psi_{\1}\\
u_{\1}\sin2\psi_{\1}\\
\rho_{\1}v_{\1}\cos2\varphi_{\1}
\end{array}
\right)\,\,.
\end{equation}
For incident P waves upont a solid/liquid interface, we only have two incidence regions
\begin{equation}
 0\,<\,\psi_{\1}^{^{[\mathrm{a}]}}\,<\,\arcsin \frac{v_{\1}}{v_{\2}}\,<\,\psi_{\1}^{^{[\mathrm{b}]}}\,<\,\frac{\pi}{2}\,\,.
\end{equation}
Indeed, due to the fact that $v_{\1}>u_{\1}$, the angle of the reflected S wave will always be real. In the first region also the angle of the transmitted P wave is real and consequently the reflection and transmission coefficients are both real and no lateral displacement occurs. In the second region, $\cos\psi_{\2}=i\,|\cos\psi_{\2}|$, and additional phases are found in the amplitudes of the reflected P and S
waves.

The critical angles for an incident P wave in the solid/liquid examples used in this paper are given in Table 3.

\subsection{Incidence region before the critical angle}
In this incidence region the Knott coefficients are given in terms of real reflections and transmission coefficients,

\begin{eqnarray}
E^{^{\SoLiP}}_{_{\REF(\P)}} & = & \left[\,
\frac{\rho_{\2} v_{\2} v_{\1}\cos\psi_{\1} + \rho_{\1}\cos\psi_{\2}\,(\,u_{\1}^{\2}\sin2\psi_{\1}\sin2\varphi_{\1} - v_{\1}^{\2}\cos^{\2}2\varphi_{\1}\,)}
 {\rho_{\2} v_{\2} v_{\1}\cos\psi_{\1} + \rho_{\1}\cos\psi_{\2}\,(\,u_{\1}^{\2}\sin2\psi_{\1}\sin2\varphi_{\1} + v_{\1}^{\2}\cos^{\2}2\varphi_{\1}\,)}
\,\right]^{^{2}}\,\,    \nonumber\\
E^{^{\SoLiP}}_{_{\REF(\S)}} & = &  \frac{u_{\1}\cos\varphi_{\1}}{v_{\1}\,\cos\psi_{\1}}\,\left[\,
\frac{2\,\rho_{\1}u_{\1}v_{\1}\sin2\psi_{\1}\cos2\varphi_{\1}\cos\psi_{\2}}
{\rho_{\2} v_{\2}v_{\1}\cos\psi_{\1} + \rho_{\1}\cos\psi_{\2}\,(\,u_{\1}^{\2}\sin2\psi_{\1}\sin2\varphi_{\1} + v_{\1}^{\2}\cos^{\2}2\varphi_{\1}\,)}
\,\right]^{^{2}}\,\,, \\
E^{^{\SoLiP}}_{_{\TRA(\P)}} & = & \frac{\rho_{\2}v_{\2}\cos\psi_{\2}}{\rho_{\1}v_{\1}\cos\psi_{\1}}\,\left[\,
\frac{2\,\rho_{\1}v_{\1}^{\2}\cos\psi_{\1}\cos2\varphi_{\1}}
{\rho_{\2} v_{\2}v_{\1}\cos\psi_{\1} + \rho_{\1}\cos\psi_{\2}\,(\,u_{\1}^{\2}\sin2\psi_{\1}\sin2\varphi_{\1} + v_{\1}^{\2}\cos^{\2}2\varphi_{\1}\,)}
\,\right]^{^{2}}\,\,,  \nonumber
\end{eqnarray}
and the energy conservation guaranteed by
\[
E^{^{\SoLiP}}_{_{\REF(\P)}} + E^{^{\SoLiP}}_{_{\REF(\S)}} + E^{^{\SoLiP}}_{_{\TRA(\P)}} =1\,\,.
\]
This scenario is illustrated in the plots of Fig.\,\ref{C} (c-f) for wet sand/water and granite/water for all incidence angles. For the case of vegetal soil/water, we find three travelling waves for incidence angles before the critical angle $27.82^{^{o}}$, see Fig.\,\ref{C} (a-b).

\subsection{Incidence region after the critical angle}
For incidence angles greater than the critical one, we have evanescent transmitted P waves and the Knott coefficients become
\begin{equation}
\begin{array}{rcl}
E^{^{\SoLiP}}_{_{\REF(\P)}} & = & \displaystyle{\left|\,
\frac{\rho_{\2} v_{\2} v_{\1}\cos\psi_{\1} + i\,\rho_{\1}|\cos\psi_{\2}|\,(\,u_{\1}^{\2}\sin2\psi_{\1}\sin2\varphi_{\1} - v_{\1}^{\2}\cos^{\2}2\varphi_{\1}\,)}
 {\rho_{\2} v_{\2} v_{\1}\cos\psi_{\1} + i\,\rho_{\1}|\cos\psi_{\2}|\,(\,u_{\1}^{\2}\sin2\psi_{|1}\sin2\varphi_{\1} + v_{\1}^{\2}\cos^{\2}2\varphi_{|1}\,)}
\,\right|^{^{2}}}\,\,,\\
E^{^{\SoLiP}}_{_{\REF(\S)}} & = &  \displaystyle{\frac{u_{\1}\cos\varphi_{\1}}{v_{\1}\,\cos\psi_{\1}}\,\left|\,
\frac{2\,i\,\rho_{\1}u_{\1}v_{\1}\sin2\psi_{\1}\cos2\varphi_{\1}|\cos\psi_{\2}|}
{\rho_{\2} v_{\2}v_{\1}\cos\psi_{\1} + i\,\rho_{\1}|\cos\psi_{\2}|\,(\,u_{\1}^{\2}\sin2\psi_{\1}\sin2\varphi_{\1} + v_{\1}^{\2}\cos^{\2}2\varphi_{\1}\,)}
\,\right|^{^{2}}}\,\,,
\end{array}
\end{equation}
and satisfy
\[
E^{^{\SoLiP}}_{_{\REF(\P)}} + E^{^{\SoLiP}}_{_{\REF(\S)}}  =1\,\,.
\]
In this case, the P and S reflected wave gain respectively the following phases
\begin{equation}
\begin{array}{rcl}
\alpha_{_{\GH}}^{^{\SoLiP}} &=& \displaystyle{\arctan\left[\,   \frac{
\rho_{\1}|\cos\psi_{\2}|\,(\,u_{\1}^{\2}\sin2\psi_{\1}\sin2\varphi_{\1} - v_{\1}^{\2}\cos^{\2}2\varphi_{\1}\,)}{\rho_{\2} v_{\2} v_{\1}\cos\psi_{\1}}\, \right]}\, -\\ \\
 & & \displaystyle{\arctan\left[\,   \frac{
\rho_{\1}|\cos\psi_{\2}|\,(\,u_{\1}^{\2}\sin2\psi_{\1}\sin2\varphi_{\1} + v_{\1}^{\2}\cos^{\2}2\varphi_{\1}\,)}{\rho_{\2} v_{\2} v_{\1}\cos\psi_{\1}}\, \right]}\,\,,
\end{array}
\end{equation}
and
\begin{equation}
\begin{array}{rcl}
\beta_{_{\GH}}^{^{\SoLiP}} &=& \displaystyle{\arctan\left[\,   \frac{
\rho_{\1}|\cos\psi_{\2}|\,(\,u_{\1}^{\2}\sin2\psi_{\1}\sin2\varphi_{\1} + v_{\1}^{\2}\cos^{\2}2\varphi_{\1}\,)}{\rho_{\2} v_{\2} v_{\1}\cos\psi_{\1}}\, \right]\,+\,\frac{\pi}{2}}\,\,.
\end{array}
\end{equation}

\section{Liquid-Solid interface}
In this scenario, we can only have incident and reflected $P$ waves. Consequently, we have to set $\alpha=1$, $\beta=0$, and  $u_{\1}=\varphi_{\1}=0$ in the Zoeppritz matrix equation (\ref{matuni}). The Knott energy coefficients reduce to
\begin{equation*}
\{\,E_{_{\REF(\P)}}\,,\,E_{_{\REF(\S)}}\,,\,E_{_{\TRA(\P)}} \,,\,E_{_{\TRA(\S)}}\,\} =
\left\{\,\left|\,R_{_{\P\P}}\,\right|^{^{2}}\,,\,0\,,\,
\displaystyle{\frac{\rho_{\2}v_{\2}\cos\psi_{\2}}{\rho_{\1}v_{\1}\cos\psi_{\1}}}\,\left|\,T_{_{\P\P}}\,\right|^{^{2}}\,
,\,
\displaystyle{\frac{\rho_{\2}u_{\2}\cos\varphi_{\2}}{\rho_{\1}v_{\1}\cos\psi_{\1}}}\,\left|\,T_{_{\P\S}}\,\right|^{^{2}}
\,\right\}
\end{equation*}
and the reflection and transmission coefficients will be determined by solving the Zoeppritz matrix equation

\begin{equation}
\left(\,
\begin{array}{rrrr}
-\,\sin\psi_{\1} & -\,1 &\sin\psi_{\2} &-\,\cos\varphi_{\2}\\
\cos\psi_{\1} & 0 & \cos\psi_{\2} & \sin\varphi_{\2} \\
0 &  0& u_{\2}\sin2\psi_{\2} & -\,v_{\2}\cos2\varphi_{\2}\\
-\,\rho_{\1}v_{\1} & 0 & \rho_{\2}v_{\2}\cos2\varphi_{\2} & \rho_{\2}u_{\2}\sin2\varphi_{\2}
\end{array}
\,\right)\,\,\left(\,\begin{array}{c}
R_{_{\P\P}}\\
R_{_{\P\S}}\\
T_{_{\P\P}}\\
T_{_{\P\S}}
\end{array}
\right) =  \left(\,\begin{array}{c}
\sin\psi_{\1}\\
\cos\psi_{\1}\\
0\\
\rho_{\1}v_{\1}
\end{array}
\right)\,\,.
\end{equation}
For a liquid/solid interface, we find three distinct incidence subregions
\begin{equation}
0\,<\,\psi_{\1}^{^{[\mathrm{I}]}}\,<\,\arcsin \frac{v_{\1}}{v_{\2}}\,<\,\,\psi_{\1}^{^{[\mathrm{II}]}}\,<\,\arcsin \frac{v_{\1}}{u_{\2}} \,<\,\psi_{\1}^{^{[\mathrm{III}]}}\,<\,\frac{\pi}{2}\,\,.
\end{equation}
The critical angles for an incident P wave in the liquid/solid examples used in this paper are given in Table 4.

\subsection{First incidence region}
In the first incidence region $\cos\psi_{\2}$ and $\cos\varphi_{\2}$, are both real and consequently we have a
travelling (back) reflected P wave and
two travelling (forward) transmitted P and S waves with Knott energy coefficients given by
\begin{eqnarray}
E^{^{\LiSo}}_{_{\REF(\P)}} & = & \left[\,
\frac{\rho_{\2}\cos\psi_{\1}(\,v_{\2}^{\2}\cos^{\2}2\varphi_{\2} + u_{\2}^{\2}\sin2\psi_{\2}\sin2\varphi_{\2}\,) - \rho_{\1}v_{\1}v_{\2}\cos\psi_{\2} }{\rho_{\2}\cos\psi_{\1}(\,v_{\2}^{\2}\cos^{\2}2\varphi_{\2} + u_{\2}^{\2}\sin2\psi_{\2}\sin2\varphi_{\2}\,) + \rho_{\1}v{\1}v_{\2}\cos\psi_{\2}   }
\,\right]^{^{2}}    \nonumber\\
E^{^{\LiSo}}_{_{\TRA(\P)}} & = & \frac{\rho_{\2}v_{\2}\cos\psi_{\2}}{\rho_{\1}v_{\1}\cos\psi_{\1}}\,\left[\,
\frac{2\,\rho_{\1}v{\1}v_{\2}\cos\psi_{\1}\cos2\varphi_{\2}}
{\rho_{\2}\cos\psi_{\1}(\,v_{\2}^{\2}\cos^{\2}2\varphi_{\2} + u_{\2}^{\2}\sin2\psi_{\2}\sin2\varphi_{\2}\,) + \rho_{\1}v{\1}v_{\2}\cos\psi_{\2}}
\,\right]^{^{2}}    \\
E^{^{\LiSo}}_{_{\TRA(\S)}} & = &  \frac{\rho_{\2}u_{\2}\cos\varphi_{\2}}{\rho_{\1}v{\1}\cos\psi_{\1}}\,\left[\,
\frac{2\,\rho_{\1}v_{\1}u_{\2}\cos\psi_{\1}\sin2\psi_{\2}}
{\rho_{\2}\cos\psi_{\1}(\,v_{\2}^{\2}\cos^{\2}2\varphi_{\2} + u_{\2}^{\2}\sin2\psi_{\2}\sin2\varphi_{\2}\,) + \rho_{\1}v{\1}v_{\2}\cos\psi_{\2}}
\,\right]^{^{2}} \nonumber
\end{eqnarray}
and satisfying
\[
E^{^{\LiSo}}_{_{\REF(\P)}} + E^{^{\LiSo}}_{_{\TRA(\P)}} + E^{^{\LiSo}}_{_{\TRA(\S)}} =1\,\,.
\]
This is for example the case of water/vegetal soil for all the incidence angles, see Fig.\,\ref{D}(a-b).
For water/wet sand and water/granite interfaces this scenario is respectively seen before the critical angles
$48.59^{^{o}}$, Fig.\,\ref{D} (c-d), and $14.48^{^{o}}$, see Fig.\,\ref{D} (e-f).

\subsection{Second incidence region}
This is the incidence region for which only the transmitted P angle is imaginary,  $\cos\psi_{\2}=i\,|\cos\psi_{\2}|$. In this case, we have an evanescent transmitted P and a travelling transmitted S wave. the Knott coefficients are given by
\begin{equation}
\begin{array}{rcl}
E^{^{\LiSo}}_{_{\REF(\P)}} & = & \displaystyle{\left|\,
\frac{\rho_{\2}\cos\psi_{\1}(\,v_{\2}^{\2}\cos^{\2}2\varphi_{\2} + \,i\,u_{\2}^{\2}|\sin2\psi_{\2}|\sin2\varphi_{\2}\,) -\,i\, \rho_{\1}v_{\1}v_{\2}|\cos\psi_{\2}| }{\rho_{\2}\cos\psi_{\1}(\,v_{\2}^{\2}\cos^{\2}2\varphi_{\2} + \,i\, u_{\2}^{\2}|\sin2\psi_{\2}|\sin2\varphi_{\2}\,) + \,i\,\rho_{\1}v{\1}v_{\2}|\cos\psi_{\2}| }
\,\right|^{^{2}}}\,\,,   \\ \\
E^{^{\LiSo}}_{_{\TRA(\S)}} & = &  \displaystyle{\frac{\rho_{\2}u_{\2}\cos\varphi_{\2}}{\rho_{\1}v_{\1}\cos\psi_{\1}}\,\left|\,
\frac{2\,i\,\rho_{\1}v_{\1}u_{\2}\cos\psi_{\1}|\sin2\psi_{\2}|}
{\rho_{\2}\cos\psi_{\1}(\,v_{\2}^{\2}\cos^{\2}2\varphi_{\2} + \,i\, u_{\2}^{\2}|\sin2\psi_{\2}|\sin2\varphi_{\2}\,) + \,i\,\rho_{\1}v{\1}v_{\2}|\cos\psi_{\2}| }
\,\right|^{^{2}} }\,\,,
\end{array}
\end{equation}
and they guarantee the energy conservation
\[
E^{^{\LiSo}}_{_{\REF(\P)}} + E^{^{\LiSo}}_{_{\TRA(\S)}} =1\,\,.
\]
The reflected P wave becomes complex and gains the following GH phase

\begin{equation}
\begin{array}{rcl}
\alpha_{_{\GH}}^{^{\LiSo}} &=& \displaystyle{\arctan\left(\,   \frac{\rho_{\2}\cos\psi_{\1}u_{\2}^{\2}\,|\sin2\psi_{\2}|\,\sin2\varphi_{\2} - \rho_{\1}v_{\1}v_{\2}|\cos\psi_{\2}|}{\rho_{\2}\cos\psi_{\1}v_{\2}^{\2}\cos^{\2}2\varphi_{\2}}\, \right)} -\\ \\
 & & \displaystyle{\arctan\left(\,   \frac{\rho_{\2}\cos\psi_{\1}u_{\2}^{\2}\,|\sin2\psi_{\2}|\,\sin2\varphi_{\2} + \rho_{\1}v_{\1}v_{\2}|\cos\psi_{\2}|}{\rho_{\2}\cos\psi_{\1}v_{\2}^{\2}\cos^{\2}2\varphi_{\2}} \,\right)}\,\,.
\end{array}
\end{equation}
This is for example the case illustrated in Fig.\,\ref{D} (c-d) after the critical angle $48.59^{^{o}}$ (water/wet sand)
and in Fig.\,\ref{D} (e-f) between in $14.48{^{o}}$ and  $27.04^{^{o}}$ (water/granite).

\subsection{Third incidence region}
In the last incidence subregion both the transmitted P and S waves are evanescent and the Knott coefficient for the reflected P wave becomes
\begin{eqnarray}
E^{^{\LiSo}}_{_{\REF(\P)}} & = & \left|\,
\frac{\rho_{\2}\cos\psi_{\1}(\,v_{\2}^{\2}\cos^{\2}2\varphi_{\2} - u_{\2}^{\2}|\sin2\psi_{\2}\sin2\varphi_{\2}|\,) -\,i\, \rho_{\1}v_{\1}v_{\2}|\cos\psi_{\2}| }{\rho_{\2}\cos\psi_{\1}(\,v_{\2}^{\2}\cos^{\2}2\varphi_{\2} - u_{\2}^{\2}|\sin2\psi_{\2}\sin2\varphi_{\2}|\,) + \,i\,\rho_{\1}v_{\1}v_{\2}|\cos\psi_{\2}|   }
\,\right|^{^{2}}=1\,\,.
\end{eqnarray}
The reflection coefficient for the reflected P wave is then given by $\exp\{\,i\,\beta_{_{\GH}}^{^{\LiSo}}\,\}$  with
\begin{equation}
\beta_{_{\GH}}^{^{\LiSo}} =  -\,2\,\arctan\left[\,  \frac{\rho_{\1}v_{\1}v_{\2}|\cos\psi_{\2}|}{\rho_{\2}\cos\psi_{\1}(\,v_{\2}^{\2}\cos^{\2}2\varphi_{\2} - u_{\2}^{\2}|\sin2\psi_{\2}\sin2\varphi_{\2}|\,)}   \right]\,\,.
\end{equation}
This scenario is for example seen in Fig.\, \ref{D}(e-f) for incidence angles greater than $27.04^{^{o}}$ (water/granite interface).

\section{Lateral displacements}
In 1948\cite{ART}, Artmann used the method of stationary phase and, analyzing the additional phase of the Fresnel coefficients describing the total reflection of optical waves, theoretically explained  the lateral displacement for transverse electric (TE) waves observed, one year before, by Goos and H\"anchen. He also predicted a different displacement for transverse magnetic (TM) waves, experimentally confirmed in 1949\cite{GH2}.

In this section, we briefly introduce the Artmann analytical tool to find the lateral shift of optical waves and then apply it to find the lateral displacements of reflected seismic P and S waves. In doing it, let us introduce a real angle distribution, $g(\theta,\theta_{\0})$,  centered in $\theta_{\0}$, the angle of incidence of an optical beam upon a dielectric/air interface. This angular distribution allows to determine the behavior of the
incident and reflected wave packets  through the integrals
\[{\mathrm{incident\,\,wave}} \,\,:\,\, \int {\mathrm{d}}\theta\,g(\theta,\theta_{\0})\,\exp[\,i\,k\,(\,\sin\theta\,y\,+\,\cos\theta\,z\,\,)\,]\]
and
\[{\mathrm{reflected\,\,wave}} \,\,:\,\, \int {\mathrm{d}}\theta\,g(\theta,\theta_{\0})\,R(\theta)\,\exp[\,i\,k\,(\,\sin\theta\,y\,-\,\cos\theta\,z\,\,)\,]\,\,,\]
where $R(\theta)$ is the Fresnel coefficient for the reflected wave obtained by solving the Maxwell equations
and imposing the field continuity conditions. The stationary phase method idea is used to estimate the beam propagation without solving the integrals. This can be done by observing that asymptotically, when the phase is large enough to generate rapid oscillations, the integrand contributions cancel out,  except at stationary points, i.e. the points where the phase derivative is null. For the incident wave this happens for
\[\left[\partial_{\theta} (\,k\,\sin\theta\,y\,+\,k\,\cos\theta\,z\,)\right]_{\0}=0
\,\,\,\,\,\,\,
\Rightarrow
\,\,\,\,\,\,\,y_{\mathrm{inc}}=\tan\theta_{\0}\,z\,\,.
 \]
When the reflection coefficient is complex, $|R(\theta)|\,\exp[\,i\,\Phi_{_{\mathrm{GM}}}(\theta)]$, the spatial phase of the reflected wave gains an additional phase. By using the method of stationary phase, we find
\[\left\{\partial_{\theta} [\,k\,\sin\theta\,y\,-\,k\,\cos\theta\,z\,+\,\Phi_{_{\mathrm{GH}}}(\theta)\,]\right\}_{\0}=0
\,\,\,\,\,\,\,
\Rightarrow
\,\,\,\,\,\,\,y_{\mathrm{ref}}=-\,\tan\theta_{\0}\,z \,-\,\Phi^{^{'}}_{_{\mathrm{GH}}}(\theta_{\0})/\,k\,\cos\theta_{\0}\,\,.
 \]
Observing that $k=2\pi/\lambda$, the lateral displacement of the reflected optical beam is proportional to $\lambda$. Due to the fact that the Fresnel coefficients are different from TE and TM, the lateral shift also depends on the light polarization.

In the case of seismic waves incident on a solid/liquid interface, we have, in general, two reflected waves and consequently we have to distinguish between two lateral displacements, the Compressional or Primary GH (PGH) shift
\begin{equation}
y_{\mathrm{ref}}^{^{[\mathrm{P}]}}=-\,\tan\psi_{\1}\,z \,-\,\frac{v_{\1}}{\omega\,\cos\psi_{\1}}\,\frac{\partial \Phi_{_{
\mathrm{GH}}}}{\partial \psi_{\1}}
\end{equation}
and the  Shear or Secondary  GH (SGH) shift
\begin{equation}
y_{\mathrm{ref}}^{^{[\mathrm{S}]}}=-\,\tan\varphi_{\1}\,z \,-\,\frac{u_{\1}}{\omega\,\cos\psi_{\1}}\,\frac{\partial \Phi_{_{\mathrm{ GH}}}}{\partial \varphi_{\1}}\,\,.
\end{equation}
Lateral displacements for both the P and S reflected waves are found for incoming S waves in the following incidence region
\[\arcsin\frac{u_{\1}}{v_{\2}}\,<\,\varphi_{\1}\,<\,\arcsin\frac{u_{\1}}{v_{\1}}\]
and for incoming P waves for
\[\arcsin\frac{v_{\1}}{v_{\2}}\,<\,\psi_{\1}\,<\,\frac{\pi}{2}\,\,.\]
This happens when the liquid P wave velocity is greater than the velocity of the P (and consequently S) wave  propagating in the solid. For the examples examined in this paper, this only occurs for the vegetal soil/water scenario. We find lateral displacements for both the P and S reflected waves,
for incident S waves between the angle $11.54^{^o}$ and  $25.38^{^o}$, see Fig.\,\ref{E}(a), and for incident P waves after the angle $27.82^{^o}$, see Fig.\,\ref{F}(a).

In Optics, we always have total reflection for an incidence angle greater than the critical one. In the seismic solid/liquid  case, total reflection is only reached for incident S waves when
\[ {\mathrm{max}}\left[\,\arcsin\frac{u_{\1}}{v_{\1}}\,,\, \arcsin\frac{u_{\1}}{v_{\2}}\,\right]\,<\,
 \varphi_{\1}\,<\,\frac{\pi}{2}\,\,.\]
The lateral displacement for totally reflected S waves is shown in Fig.\,\ref{E} for vegetal soil/water after $25.38^{^o}$ (a) and for wet sand/water after  $23.58^{^o}$ (b).

For the liquid/solid scenario, the incidence angle condition which guarantees total reflection for P waves is
\[\arcsin\frac{v_{\1}}{u_{\2}}\,<\,\psi_{\1}\,<\,\arcsin\frac{\pi}{2}\,\,.\]
This means that the propagation velocity of S waves in the solid has to be greater than the velocity of the P waves propagating in the liquid. This, for example, happens for the water/granite case when in incidence angle is greater than  $27.04^{^o}$, see Fig.\,\ref{F}(c).

It is interesting to observe that we have total reflection when the following two condition are simultaneously satisfied: two critical angles appear and the incidence angle is in the third incidence region, i.e. in the region where the incidence angle is  greater than the second critical angle, see Fig.\,\ref{E}(a-b) and Fig.\,\ref{F}(c).  In these cases, with respect to the optical case, a completely new phenomenon occurs. The presence of a maximum just after the second critical angle. In particular, this is clearly evident for the liquid/solid case where the GH phase is
\[
\beta_{_{\GH}}^{^{\LiSo}} =  -\,2\,\arctan\left[\,  \frac{\rho_{\1}v_{\1}v_{\2}|\cos\psi_{\2}|}{\rho_{\2}\cos\psi_{\1}(\,v_{\2}^{\2}\cos^{\2}2\varphi_{\2} - u_{\2}^{\2}|\sin2\psi_{\2}\sin2\varphi_{\2}|\,)}   \right]\,\,.
\]
The study of the derivative of this phase is shown in Fig.\,\ref{G} for different values of $\rho_{\2}/\rho_{\1}$, $u_{\2}/v_{\1}$, and $v_{\2}/v_{\1}$. In Fig.\,\ref{G}(a), we see that,  for equal ratios of the  solid/liquid propagation velocities, changing  density ratio  the maximum of the lateral displacement changes proportionally to the density ratio but practically at the same incidence angle. The liquid/solid  phase is of the form $\arctan[{\mathrm{num}}/\rho\,{\mathrm{den}}]$, and consequently  the lateral displacement given by $\rho\,({\mathrm{num}}'{\mathrm{den}}-{\mathrm{num}}\,{\mathrm{den}}')/({\mathrm{num}}^{^{2}}+\rho^{^{2}} {\mathrm{den}}^{^{2}})$. The plots of Fig.\,\ref{G}(a) then suggests that the incidence angle, for which a maximum shift is found, is  obtained by solving
\begin{equation}
v_{\2}^{\2}\cos^{\2}2\varphi_{\2} = u_{\2}^{\2}|\sin2\psi_{\2}\sin2\varphi_{\2}|\,\,.
\end{equation}
Observing  that
\[
\begin{array}{ll}
\displaystyle{\sin\varphi_{\2}=\frac{u_{\2}}{v_{\1}}\,\sin\psi_{\1}}\,\,\,,\,\,\,&
\displaystyle{\cos\varphi_{\2}=i\,\sqrt{\left(\frac{u_{\2}}{v_{\1}}\,\sin\psi_{\1}\right)^{^{2}}-1}}\,\,\,,\\
\displaystyle{\sin\psi_{\2}=\frac{v_{\2}}{v_{\1}}\,\sin\psi_{\1}}\,\,\,, &
\displaystyle{\cos\psi_{\2}=i\,\sqrt{\left(\frac{v_{\2}}{v_{\1}}\,\sin\psi_{\1}\right)^{^{2}}-1}}\,\,\,,
\end{array}
\]
the previous equation can be rewritten as follows
\begin{equation}
v_{\2}^{\2}\left[\,1-2\,\left(\,\frac{u_{\2}}{v_{\1}}\,\sin\psi_{\1}\,\right)^{^{2}}\right]^{^{2}} = 4\, \frac{u_{\2}^{^{3}}v_{\2}}{v_{\1}^{^{2}}}\,   \sin^{\2}\psi_{\1}\,\sqrt{\left(\frac{u_{\2}}{v_{\1}}\,
\sin\psi_{\1}\right)^{^{2}}-1}\sqrt{\left(\frac{v_{\2}}{v_{\1}}\,\sin\psi_{\1}\right)^{^{2}}-1}\,\,.
\end{equation}
After simple algebraic manipulations, we find a polynomial equation in the variable $x=\sin^{\2}\psi_{\1}$,
\begin{equation}
\label{PolEq}
16\,u^{\6}\,(\,v^{\2}\,-\,u^{\2}\,)\,x^{\3}\,+\,8\,u^{\4}\,(\,2\,u^{\2}\,-\,3\,v^{\2}\,)\,x^{\2}\,+\,8\,u^{\2}\,v^{\2}\, x\,-\,v^{\2}=0\,\,,
\end{equation}
where $u=u_{\2}/v_{\1}$ and $v=v_{\2}/v_{\1}$. This polynomial equation allows to calculate, in the liquid/solid scenario, the incidence angle at which the GH lateral displacement is maximized. For a water/granite interface,
$u=33/15$ and $v=4$, Eq.\,(\ref{PolEq}) gives a real solution at $x\approx0.2416$. Consequently, the incidence angle which maximize the GH lateral shift is found at
\[
\psi_{\1}^{^{[\mathrm{max}]}}=\arcsin\sqrt{0.2416}=29.44^{^{o}}\,\,,
\]
see Fig.\,\ref{F}(c).

The divergence at the critical angle and the discontinuity between the region before and after the critical incidence  have recently been discussed and solved, in Optics, by using the wave packet formalism\cite{ANA1,ANA2,ANA4}. It was also proven that the plane wave analysis remains correct outside the critical region
\[\left[\,\theta_{\mathrm{cri}}\,-\,\frac{\lambda}{{\mathrm{w}_{\0}}}\,,\,\theta_{\mathrm{cri}}\,+\,\frac{\lambda}{{\mathrm{w}_{\0}}} \,\right]\,\,,\]
where the phase in the integrand can be approximated by using the stationary phase method and the integral analytically solved\cite{ANA3}.  This clearly also occurs for seismic waves. This means that the plane wave analysis presented in this paper reproduces the correct results for incidence angle outside the critical regions.

Of particular interest are the cases in which total reflection for a single wave occurs. In such cases, we find a maximum after the second critical angle and this effect is amplified in the liquid/solid scenario. As observed before, the plane wave analysis is valid outside the critical region. The maximum obtained in the plane wave analysis is thus valid also for wave packet with a beam waist
${\mathrm{w}}_{\0}$ if critical angle, maximal angle, velocities, frequency, and beam waist
satisfy
\[
\psi_{\1}^{^{[\mathrm{max}]}}\,>\,\arcsin\frac{v_{\1}}{u_{\2}}\,+\,\frac{v_{\1}}{\omega\,{\mathrm{w}}_{\0}}\,\,.
\]
For a water/granite interface and incident waves with a frequency $10$ KHz, this implies the following constraint on the beam waist

\[{\mathrm{w}}_{\0} \,>\, \frac{0.15\,\,{\mathrm{m}}}{29.44-27.04}\,\frac{180}{\pi}\,\approx\, 3.5\,{\mathrm{m}}\,\,. \]

\section{Conclusions}

The study of the GH shift has been a continuous source of excitement in the Optics community. Since its first experimental evidence in 1947\cite{GH1}, the central interests were the theoretical understanding of the phenomenon and the possibility to find analytical expressions for predicting the lateral displacements. The first analytical formula, based on the stationary phase method, was given by Artmann in 1948\cite{ART}. He observed that when the light is totally reflected the reflection coefficient becomes complex and the additional phase is responsible for the shift.

Even being true that the plane wave approach contains divergences that can only be removed by using the wave packet formalism\cite{ANA3,ANA4}, it is important to recall that, outside the critical region, the plane wave approach gives results in full agreement  with the ones obtained by using wave packets.  In this spirit, by using plane waves and the stationary phase method, we presented a detailed analysis of lateral displacements of the reflected P and S waves in the solid/liquid and liquid/solid scenarios, confident that, as it is done in Optics, the divergences at critical angles can later be removed by treating the problem within the wave packet formalism.

The Goos-H\"anchen effect is a phenomenon of Classical Optics in which a light beam reflecting off a surface  is spatially shifted as a consequence of its brief penetration through the surface before bouncing back.  The same phenomenon occurs for acoustic waves where, due to the matrix structure of the Zoeppritz equations, in general, two critical angles are found. In seismic data, contrary to  what happens in Optics, a total reflection can occur also for real reflection coefficients. In this case, no lateral shift is observed in the reflected wave. This is for example the case of reflected S waves  in the vegetal soil/water, wet sand/water and granite/water scenarios for incidence at  $11.54^{^{o}}$,  $17.46^{^{o}}$, and  $33.37^{^{o}}$, see Fig.\,\ref{B}(a,c,e),  and of reflected P waves in the vegetal soil/water, water/wet sand, and water/granite scenarios for incidence at  $27.82^{^{o}}$,  $48.59^{^{o}}$, and  $14.48^{^{o}}$, see Fig.\,\ref{C}(a) and Fig.\,\ref{D}(c,e).

The analysis presented in this paper shows positive and negative lateral displacements for the reflected S and P waves and the presence of a local maximum shift just after the second critical angle, see for example the case of reflected S waves in the vegetal soil/water and wet sand/water scenarios for incidence greater than  $25.38^{^{o}}$  and  $23.58^{^{o}}$, see Fig.\,\ref{E}(a,b) and the case of reflected P waves in
water/granite scenario for incidence greater that  $27.04^{^{o}}$, see Fig.\,\ref{F}(c).
In this last case, the lateral displacement is huge, for example for incident P waves of a frequency of 10 KHz, we find a lateral displacement of approximatively  20.25 m (135 $\times$ 0.15 m), see Fig.\,\ref{F}(c). In seismic analysis, these maximal lateral displacements have to be included in the theoretical predictions to avoid discrepancies between experimental data and computational simulations.

The main goals of the study presented in this article are to offer
a basic formal mathematical introduction to the GH effect in seismic data analysis, to solve the Zoeppritz equations,  to explicitly give the complex phase of  the reflection coefficients in the solid/liquid and liquid/solid scenarios from which the lateral displacement can be calculated by using the stationary phase method, and finally to find the polynomial equation allowing  the prediction of  the incidence angle which maximizes the ZGH lateral displacement, see Eq.\,(\ref{PolEq}). In  obtaining such an equation, the simulations presented in Fig.\,\ref{G}, where the density and velocity ratios  were varied, played a fundamental role.

Usually the analogies between different physical systems help to gain an increased understanding of the phenomenon studied and sometimes open the door to new effects and challenges. The study of the ZGH effect for acoustic waves is an intriguing example of this. Clearly, there still exist several open questions such as a closed formula for the maximum lateral displacement, the shift analysis for critical incidence, and the breaking of symmetry near the critical region.

Deviations from geometrical optics are not restricted to lateral displacements.
Indeed, the spatial GH shift has an angular analogous effect\cite{ANG1,ANG2,ANG3}. This effect has recently been observed in optical experiment for incidence in the resonant Brewster region\cite{ANG4}, by using a transverse electric wave to decouple the polarization from the propagation dynamics of the beam\cite{ANG5}, and by the weak measurement technique\cite{ANG6}. The angular deviations of the beam axis with respect to the ray optics prediction appear in the incidence region of partial reflection in contrast with the lateral displacements which occur when the light is totally reflected.  While the spatial GH shift is essentially due to the phase of the Fresnel reflection coefficient, the angular  effect is mainly connected to the amplitude of the Fresnel reflection coefficient. Consequently, it is
the breaking of symmetry in the optical beam wave number distribution induced by the Fresnel reflection coefficient to cause the angular deviation\cite{ANG3}. Such a breaking of symmetry can also be seen for acoustic (scalar) waves when the reflection Zoeppritz coefficient rapidly changes, see Fig.\,\ref{B}-\ref{D}   and thus this effect is not restricted to the vector nature of light.  Of particular interest, it could be, for example, to examine the Rayleigh waves influence on acoustic beams at liquid-solid interfaces\cite{BT,DL} in view of the double peak reflected beam recently discussed in Optics\cite{ANG3}.

Seismic negative lateral displacements simulate the negative GH shifts found in Optcis \cite{NEG1,NEG2,NEG3} and should be addressed in terms of desturctive interference between the incident and reflection waves.

These topics  deserve further investigations within the
wave packet formalism. The authors feeling is that this work only represents a first step in this direction and the hope is that the study presented in this paper could stimulate further investigations and forthcoming articles on this subject.

\subsubsection*{Acknowledgements}
The authors sincerely thank Dr. Gabriel G. Maia for helpful mathematical discussions, numerical checking, and stimulating conversations  during the preparation of the manuscript.

\newpage

\begin{table}
\begin{center}
\begin{tabular}{|l||r|r|c|}\hline
medium & S wave velocity (m/s) & P  wave velocity (m/s) & density (g/cm$^{\3}$)\\
\hline \hline
water & 0 & 1500 & 1.0\\ \hline
vegetal soil & 300 & 700 & 2.4\\ \hline
wet sand & 600 & 2000 & 2.1 \\ \hline
granite & 3300 & 6000 & 2.7\\ \hline
\end{tabular}
\end{center}
\caption{In this table, we list the media, their densities, and the S and P wave velocities used to show  graphical representations of the results obtained from our analysis.}
\end{table}

\begin{table}
\begin{center}
\begin{tabular}{|l||c|c|}\hline
solid/liquid interface & arcsin($u_{\1}/v_{\1}$) & arcsin($u_{\1}/v_{\2}$)  \\
\hline \hline
vegetal soil/water & $25.38^{^{o}}$ & $11.54^{^{o}}$ \\ \hline
wet sand/water & $17.46^{^{o}}$ & $23.58^{^{o}}$ \\ \hline
granite/water & $33.37^{^{o}}$ & N/A  \\ \hline
\end{tabular}
\end{center}
\caption{Critical angles for S waves incident on vegetal soil/water, wet sand/water, and granite/water interfaces.}
\end{table}

\begin{table}
\begin{center}
\begin{tabular}{|l||c|}\hline
solid/liquid interface & arcsin($v_{\1}/v_{\2}$)  \\
\hline \hline
vegetal soil/water & $27.82^{^{o}}$ \\ \hline
wet sand/water & N/A \\ \hline
granite/water & N/A  \\ \hline
\end{tabular}
\end{center}
\caption{Critical angles for P waves incident on vegetal soil/water, wet sand/water, and granite/water interfaces.}
\end{table}

\begin{table}
\begin{center}
\begin{tabular}{|l||c|c|}\hline
liquid/solid interface & arcsin($v_{\1}/v_{\2}$) & arcsin($v_{\1}/u_{\2}$)  \\
\hline \hline
water/vegetal soil & N/A & N/A \\ \hline
water/wet sand & $48.59^{^{o}}$ & N/A \\ \hline
water/granite & $14.48^{^{o}}$ & $27.04^{^{o}}$  \\ \hline
\end{tabular}
\end{center}
\caption{Critical angles for P waves incident on water/vegetal soil, water/wet sand, and water/granite interfaces.}
\end{table}

\newpage

\ColumnFigure{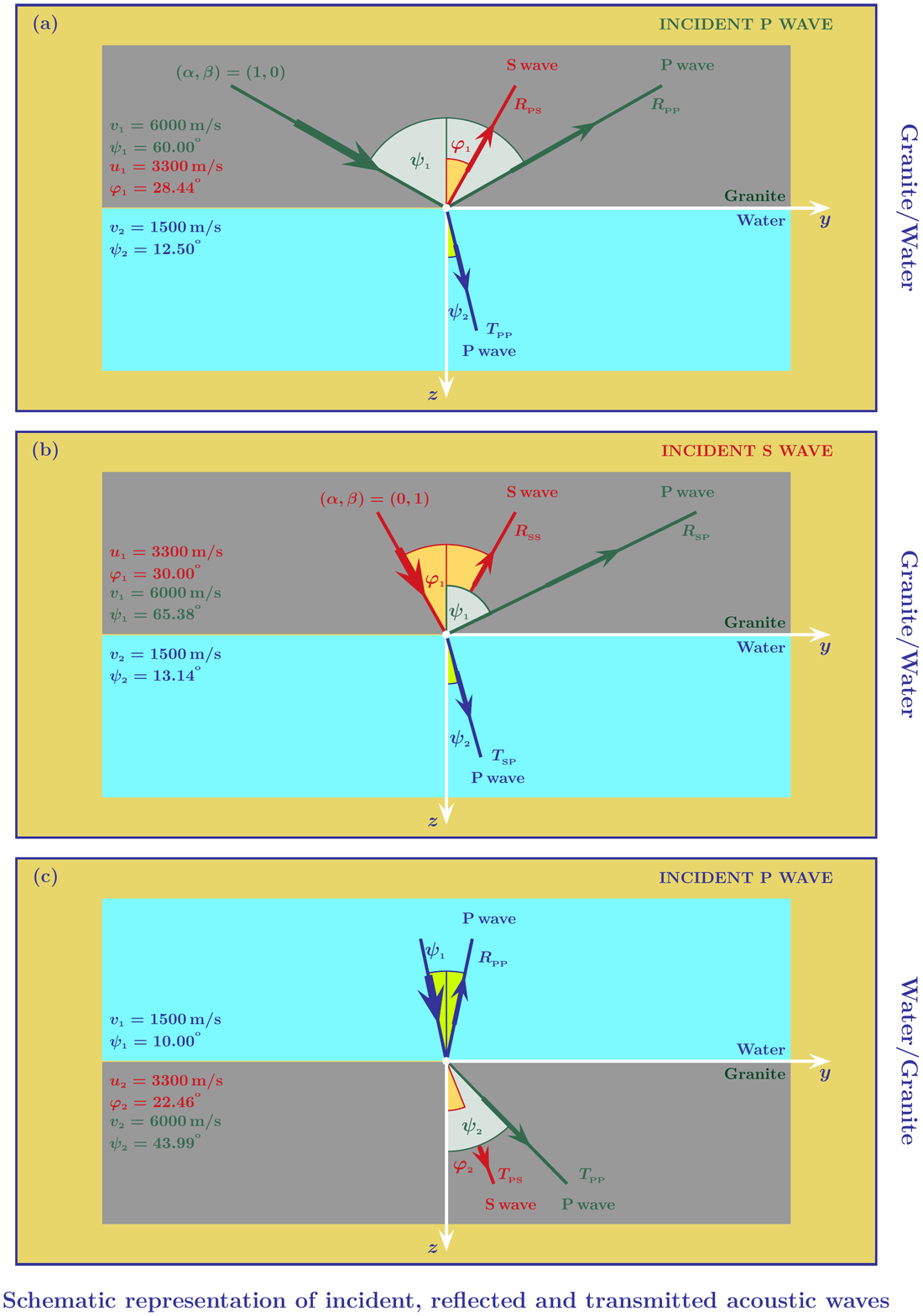}{\label{A}Reflection and transmission acoustic waves for granite/water and eater/granite interfaces. In (a), the incident P wave forms an angle of $\pi/3$ with the $z$ axis. P and S waves are reflected in granite and P waves transmitted in the water. In (b), the incident wave is an S wave and the incidence angle $\pi/6$. In (c), the incident P wave forms an angle of $\pi/18$ with the $z$ axis. P and S waves are transmitted in granite and P waves reflected in the water. The incidence angles was chosen to avoid evanescent waves.}
\ColumnFigure{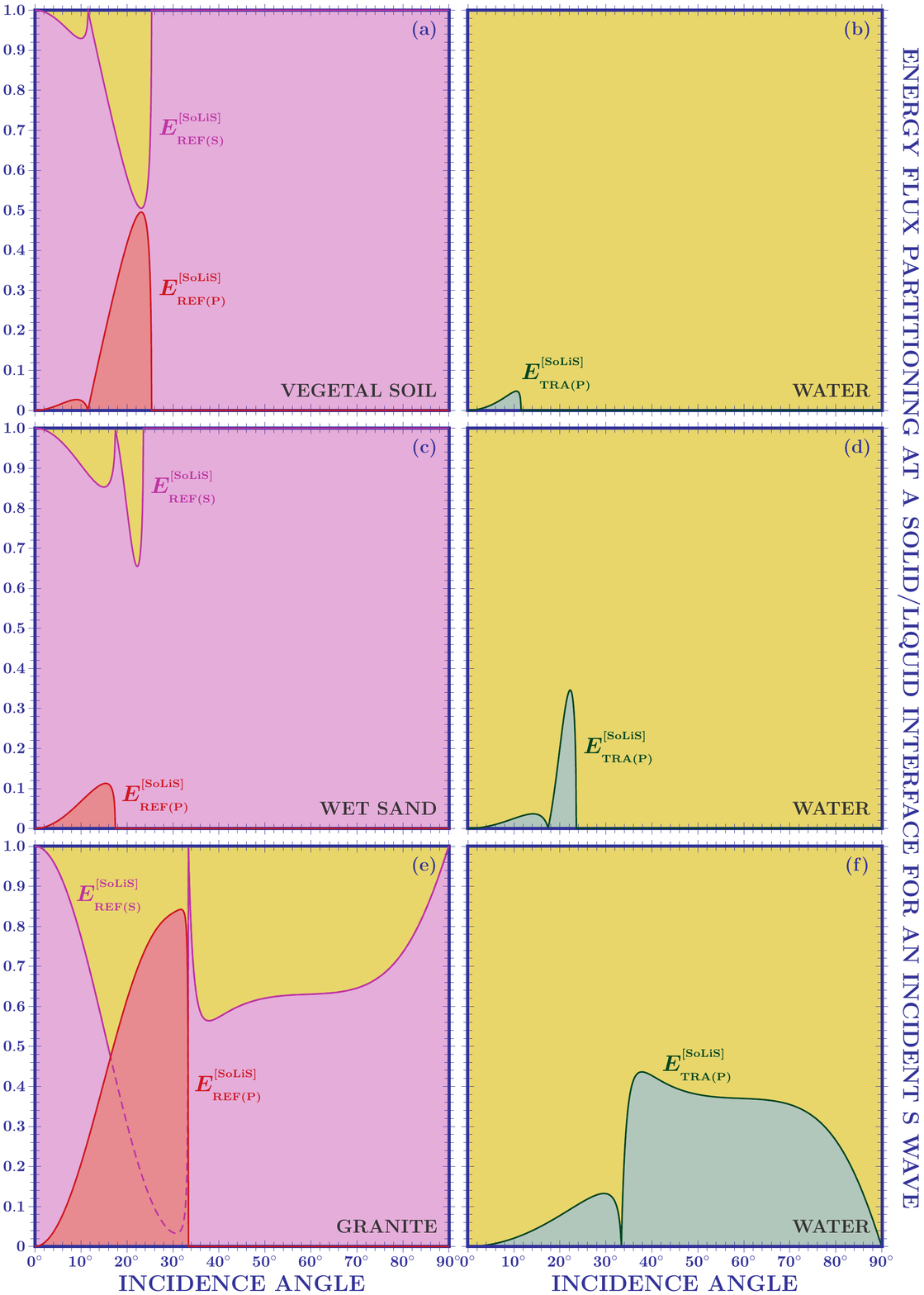}{\label{B}Energy flux partitioning at solid/liquid interfaces for an incident S wave.
Three travelling waves are present for incidence lesser than
the first critical angle, i.e. $11.54^{^{o}}$ for vegetal soil/water (a-b),  $17.46^{^{o}}$ for wet sand/water (c-d), and $33.37^{^{o}}$ for granite/water (e-f). Total reflection induced by a complex coefficient occurs  for S waves for incidence greater than the second critical angle, i.e. $25.38^{^{o}}$ for vegetal soil/water (a-b) and  $23.58^{^{o}}$ for wet sand/water.}
\ColumnFigure{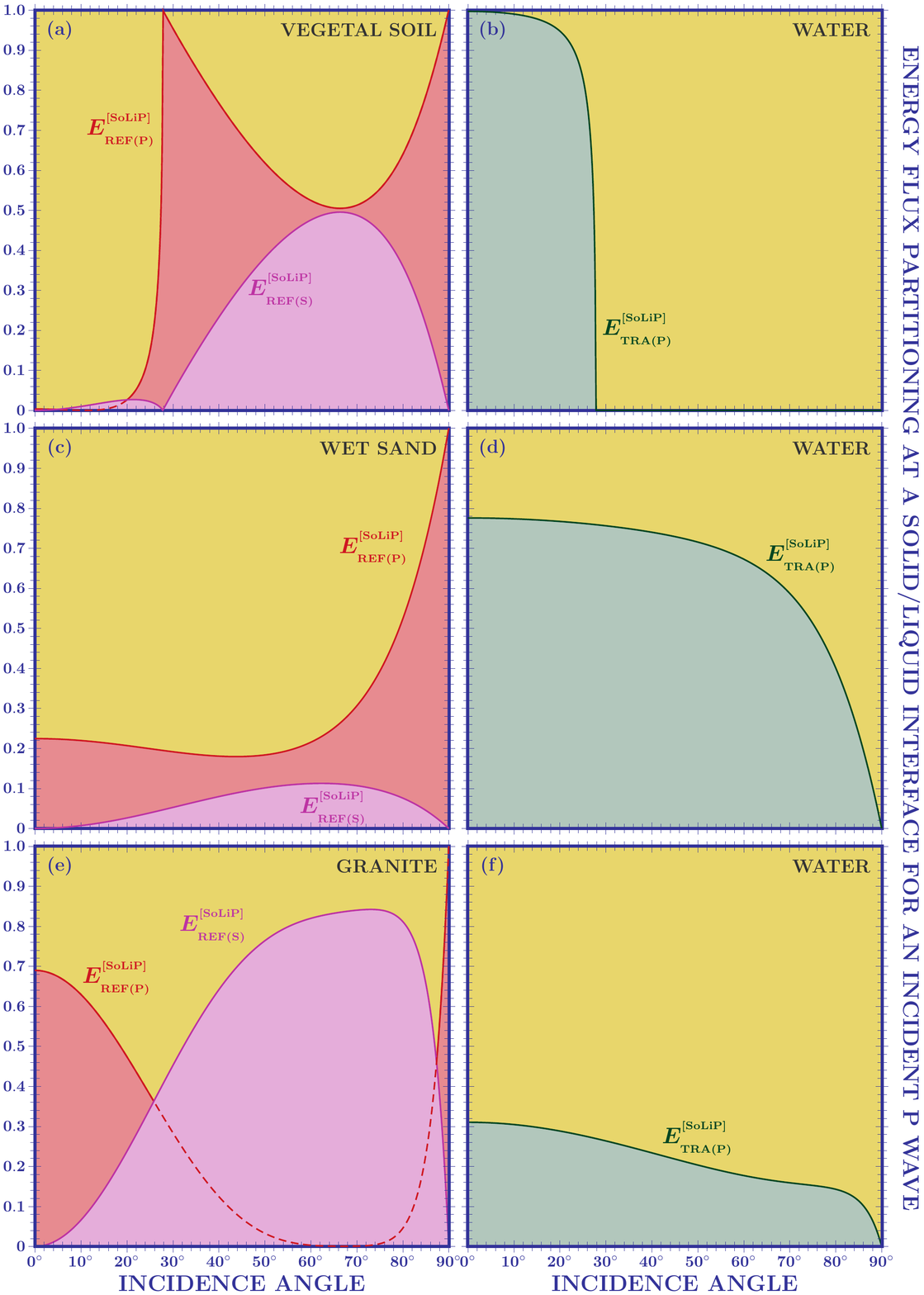}{\label{C}Energy flux partitioning at solid/liquid interfaces for an incident P wave.
For wet sand/water (c-d) and granite/water (e-f), three travelling waves are found  for all incidence angles. For vegetal soil/water (a-b) this occurs for incidence before the critical angle $27.82^{^{o}}$. Total reflection induced by a complex coefficient  is not present in these scenarios.}
\ColumnFigure{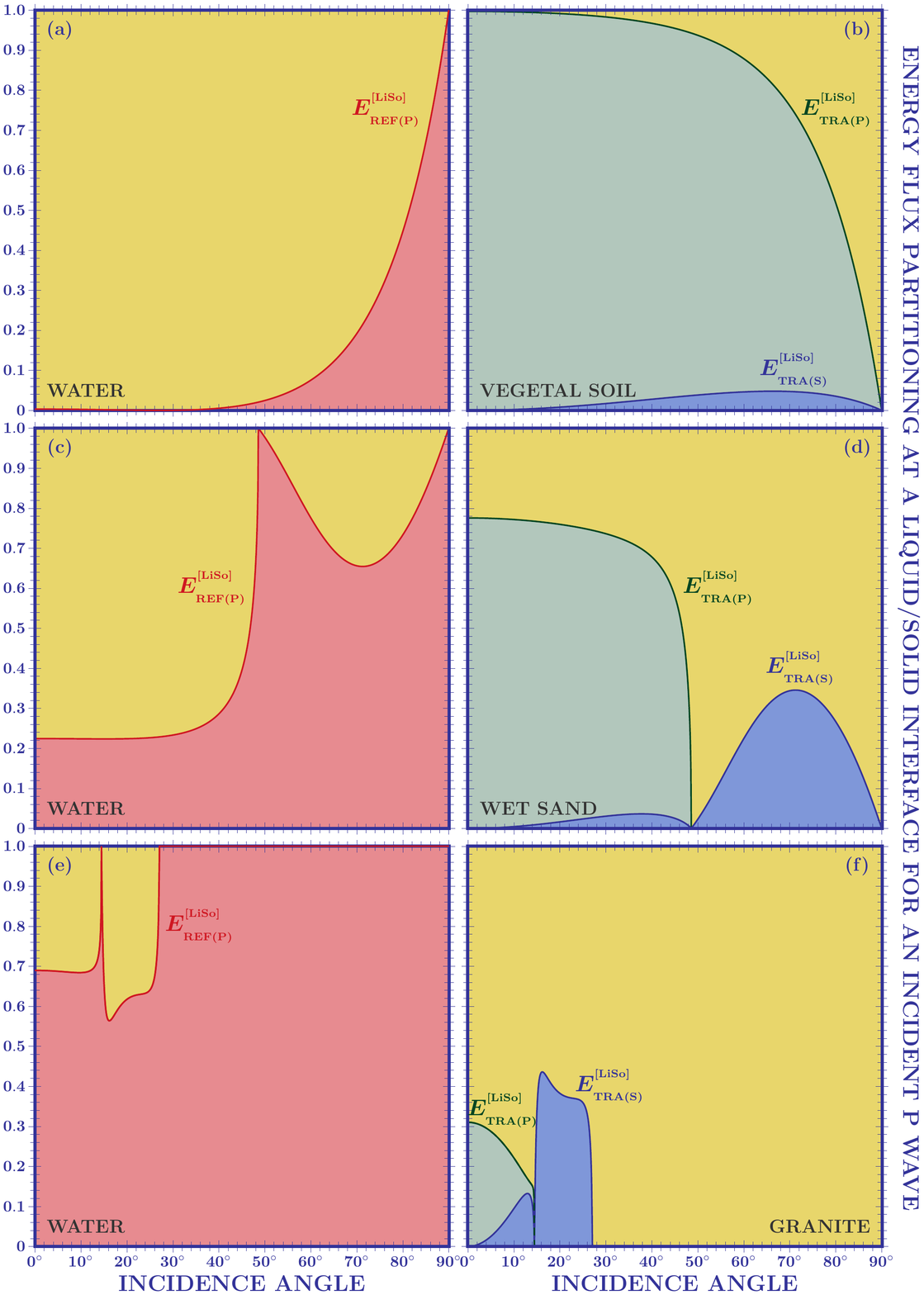}{\label{D}Energy flux partitioning at liquid/solid interfaces for an incident P wave.
Three travelling waves are found for incidence  lesser than
the first critical angle, i.e. $48.59^{^{o}}$ for wet sand/water (c-d) and  $14.48^{^{o}}$ for granite/water (e-f), and for all the incidence angles for vegetal soil/water (a-b).
Total reflection induced by a complex reflection coefficient occurs  for incidence angles greater than the second critical angle, i.e. $27.04^{^{o}}$ for granite/water (e-f).}
\ColumnFigure{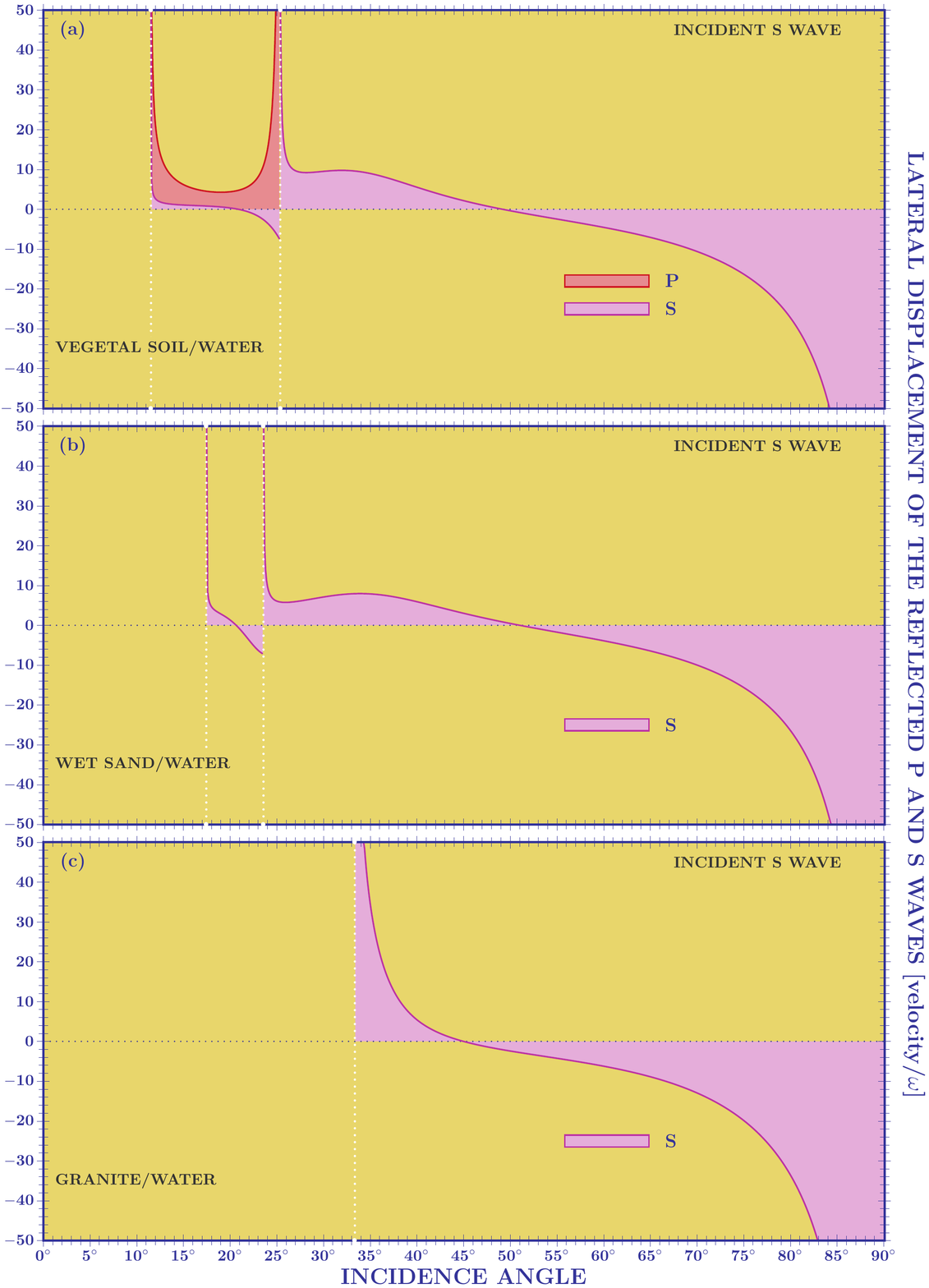}{\label{E}Lateral displacement in the solid/liquid scenarios for the reflected P and S waves in the case of an incidence S wave. In the regime of total internal reflection, after the second critical angle $25.38^{^{o}}$ for vegetal soil/water (a-b) and  after $23.58^{^{o}}$ for wet sand/water (c-d), a new maximum appears. This  behavior is not present in Optics. The divergences at critical angles are typical of the plane wave approach and they can be removed by using the wave packet formalism.}
\ColumnFigure{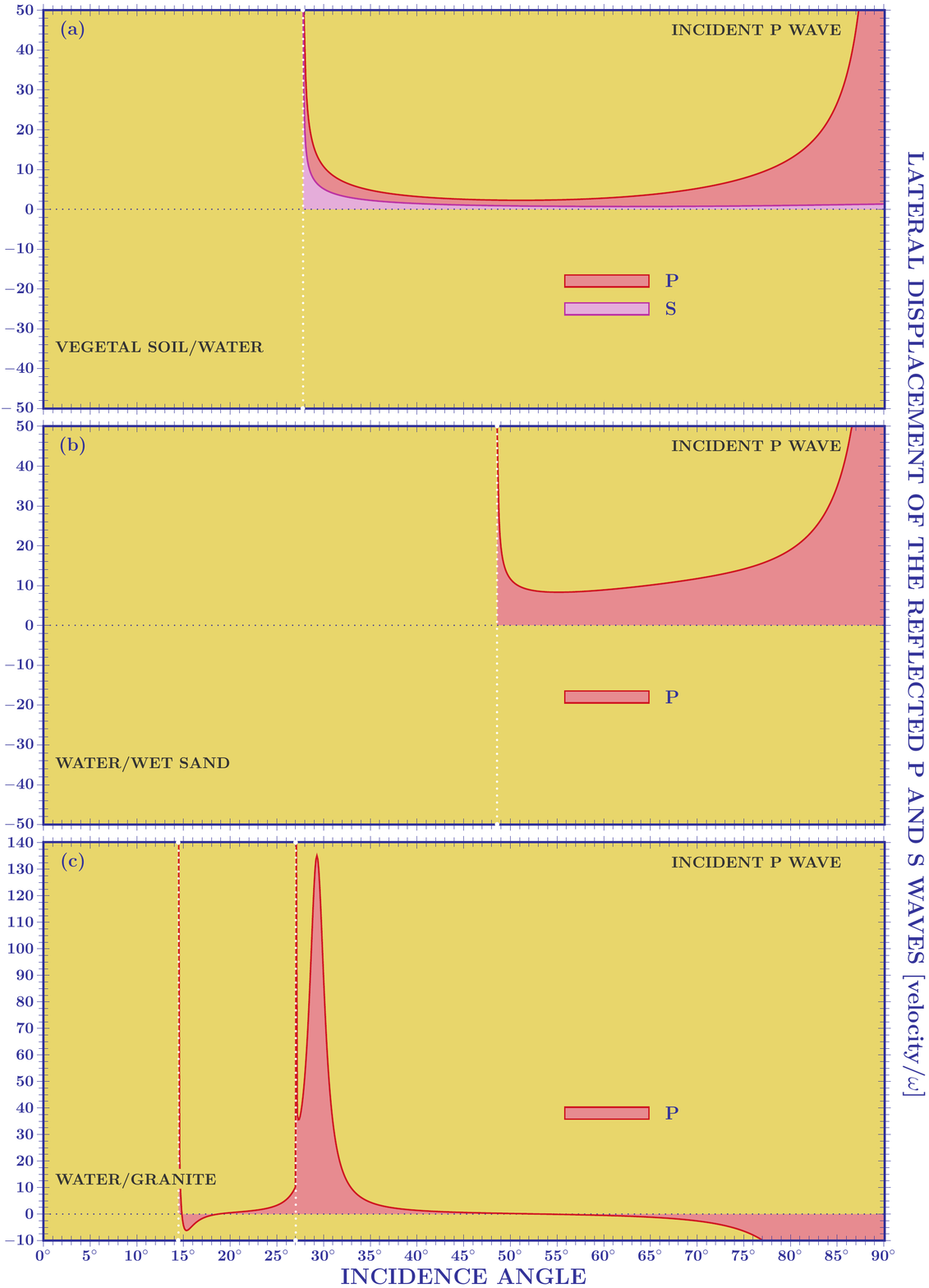}{\label{F}Lateral displacement in the solid/liquid (a) and liquid/solid (b-c) scenarios
for the reflected P and S waves in the case of an incidence P wave. In the water/granite case (c), for total internal reflection, i.e. incidence angles greater than $27.04^{^{o}}$, after the second critical angle  we observe the presence of an incidence angle which maximizes the lateral shift.}
\ColumnFigure{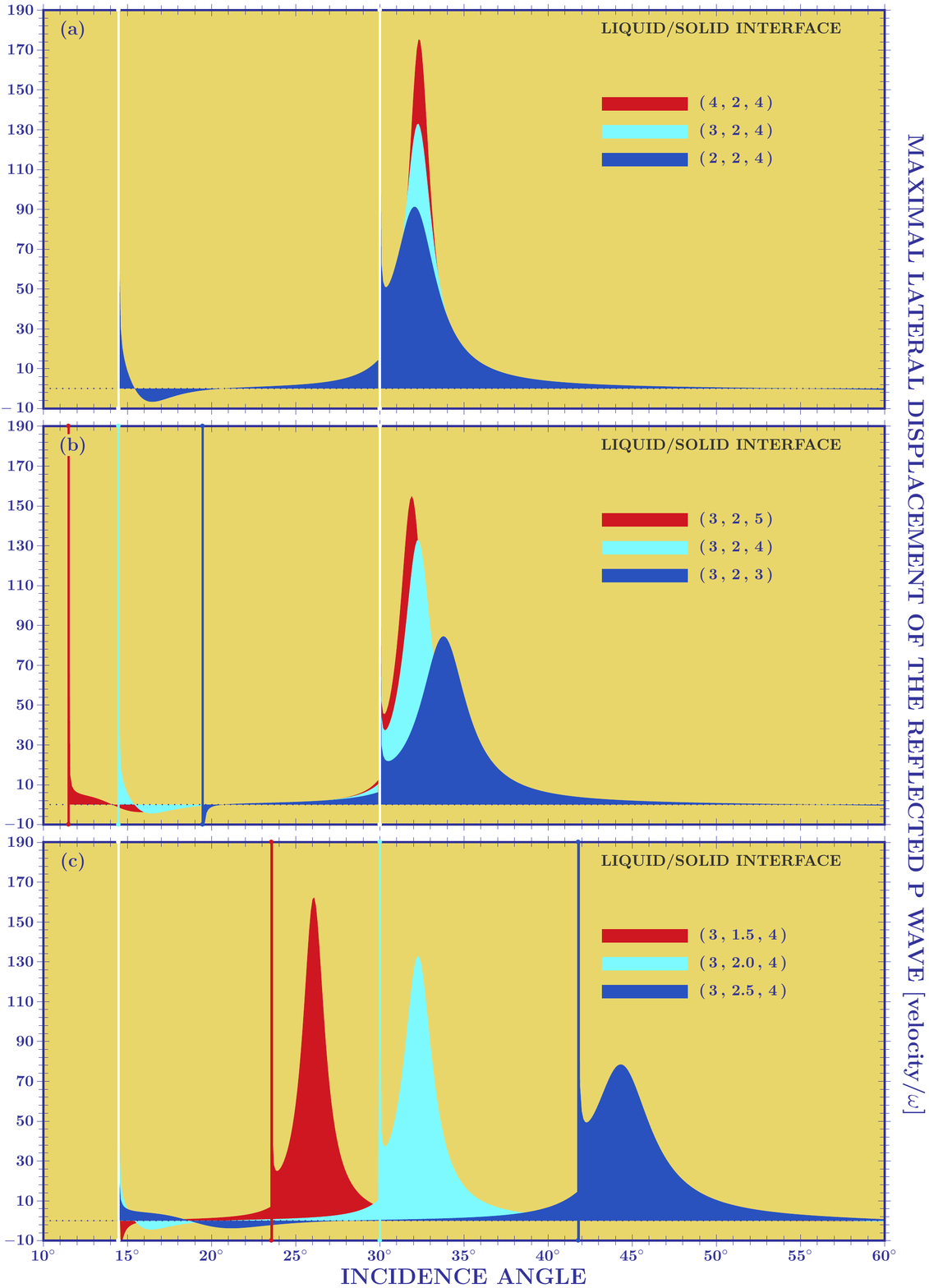}{\label{G}Lateral displacement for the liquid/solid scenario for different density and velocity ratios. In (a), the velocity ratios $u_{\2}=2\,v_{\1}$ and $v_{\2}=4\,v_{\1}$ are fixed and  the density ratio is varied $\rho_{\2}=(4,\,3\,,2)\,\rho_{\1}$. The effect of the velocity ratio variation is analyzed in (b) where $\rho_{\2}=3\,\rho_{\1}$, $u_{\2}=2\,v_{\1}$, $v_{\2}=(5,\,4\,,3)\,v_{\1}$   and in (c) where  $\rho_{\2}=3\,\rho_{\1}$, $v_{\2}=4\,v_{\1}$, $u_{\2}=(1.5,\,2.0\,,2.5)\,v_{\1}$ .}

\end{document}